\documentclass[fleqn,10pt]{wlscirep}

\usepackage{bbm}
\usepackage{amsmath,amssymb,latexsym}
\usepackage[section]{placeins}
\usepackage{epstopdf}
\usepackage{color}

\newcommand{\e}{\mathbf{e}}
\newcommand{\p}{\mathbf{p}}
\newcommand{\q}{\mathbf{q}}
\renewcommand{\r}{\mathbf{r}}
\newcommand{\x}{\mathbf{x}}

\newcommand{\N}{\mathcal{N}}

\newcommand{\U}{\mathcal{U}}

\newcommand{\micron}{\mu\mathrm{m}}
\renewcommand{\v}{\mathbf{v}}

\newcommand{\ol}[1]{{\overline{#1}}}

\newcommand{\CDF}{\mathrm{CDF}}
\newcommand{\simu}{\mathrm{sim}}

\title{
Resilience of three-dimensional sinusoidal networks in liver tissue
}

\author[1]{Jens Karschau}
\author[2]{Andre Scholich} 
\author[1,3]{Jonathan Wise} % Univ. Grenoble Alpes, CNRS, LPMMC, 38000 Grenoble, France
\author[4]{Hern{\'a}n Morales-Navarette}
\author[4]{Yannis Kalaidzidis}
\author[4,5]{Marino Zerial}
\author[1,5*]{Benjamin M Friedrich}
\affil[1]{cfaed, TU Dresden, 01069 Dresden, Germany}
\affil[2]{Max Planck Institute for the Physics of Complex Systems, 01187 Dresden, Germany}
\affil[3]{Univ. Grenoble Alpes, CNRS, LPMMC, 38000 Grenoble, France}
\affil[4]{Max Planck Institute for Molecular Cell Biology and Genetics, 01307 Dresden, Germany}
\affil[5]{Cluster of Excellence `Physics of Life', TU Dresden, 01062 Dresden, Germany}
\affil[*]{benjamin.m.friedrich@tu-dresden.de} 


\begin{abstract}
Can three-dimensional, microvasculature networks still ensure blood supply if individual links fail?
We address this question in the sinusoidal network, a plexus-like microvasculature network, 
which transports nutrient-rich blood to every hepatocyte in liver tissue,
by building on recent advances in high-resolution imaging and digital reconstruction of adult mice liver tissue.
We find that the topology of the three-dimensional sinusoidal network reflects its two design requirements 
of a space-filling network that connects all hepatocytes, while using shortest transport routes:
sinusoidal networks are sub-graphs of the Delaunay graph of their set of branching points,
and also contain the corresponding minimum spanning tree, both to good approximation.
To overcome the spatial limitations of experimental samples
and generate arbitrarily-sized networks,
we developed a network generation algorithm 
that reproduces the statistical features of 0.3-mm-sized samples of sinusoidal networks,
using multi-objective optimization for node degree and edge length distribution.
Nematic order in these simulated networks implies anisotropic transport properties,
characterized by an empirical linear relation between a nematic order parameter and the anisotropy of the permeability tensor.
Under the assumption that all sinusoid tubes have a constant and equal flow resistance, 
we predict that the distribution of currents in the network is very inhomogeneous, 
with a small number of edges carrying a substantial part of the flow. 
We quantify network resilience in terms of a permeability-at-risk, i.e.\ permeability as function of the fraction of removed edges. 
We find that sinusoidal networks are resilient to random removal of edges, but vulnerable to the removal of high-current edges.
Our findings suggest the existence of a mechanism counteracting flow inhomogeneity to balance metabolic load on the liver.
\end{abstract}
\begin{document}

\flushbottom
\maketitle

\thispagestyle{empty}

\section*{Introduction} 
Leaf venation \cite{Katifori2016}, fungal mycelium \cite{Tero2010,Alim2013},
but also
animal trails networks \cite{Perna2014}, 
river deltas \cite{Seybold2007}, and even
force networks in granular materials \cite{Radjai1996}, 
each represent natural transport networks formed by self-organization.
% NOTE: we did not add 'tree-like tracheal networks'? since it is not clear if formed by self-organization
Inside our body, the micro-vasculature forms a plexus-like, three-dimensional network of small capillaries 
that deliver nutrients to every cell in a tissue and removes waste products \cite{Kmiec2001}. 

Mathematically, transport networks are spatial networks, i.e., graphs embedded in space.
It has been proposed that the presence of cycles in these graphs provides redundancy, and thereby resilience against the failure of individual links \cite{Katifori2010}.
Yet, to the best of our knowledge, this concept has never been tested in three-dimensional, micro-vasculature networks.
Past research addressed resilience properties almost exclusively in two-dimensional biological transport networks.
It was shown that self-organization by local feedback rules can generate hierarchical networks resembling those of leaf networks, 
which optimize flow resistance upon removal of a single link \cite{Katifori2016}. 
For example, work on two-dimensional networks addressed
the balance between the cost of network formation and network resilience to random failure \cite{Bottinelli2017},
or the cost of repair after perturbations \cite{Farr2014}.

The restriction to two-dimensional networks in past research
was largely a consequence of the considerable difficulties in imaging three-dimensional networks.
Yet, topology suggests fundamental differences between two-dimensional and three-dimensional spatial networks,
as it imposes constraints on the distribution of cycles in the network \cite{Modes2016}.
Here, we take advantage of recent technological advances in high-resolution imaging of adult mouse liver tissue \cite{Morales2015,Morales2019},
allowing us to study statistical geometry and resilience of three-dimensional sinusoidal networks.

The liver is the largest metabolic organ in the human body. 
It is responsible for storage of metabolites, secretion of digestion enzymes and detoxification of blood 
\cite{Kmiec2001}. % NOTE: if more reviews needed: Elias 1949; AJ Hale: The Minute Structure of the Liver, ...
The liver is organized into millimeter-sized basic functional units termed lobules, see Figure.~\ref{figure1}A.
Each lobule comprises a central vein (CV) and a portal vein (PV) connected by a dense plexus,
termed the sinusoidal network, which transports metabolite-rich blood from PV to CV.
The sinusoidal network contacts each of the hepatocytes,
the main parenchymal cell type in the liver, which take up nutrients and toxins from the blood.
The liver serves as the central organ of blood detoxification, including the metabolism of medically relevant drugs.
Hence, an ongoing intense research effort focuses on the pharmacological modeling of the uptake of drugs from the blood stream in the liver
\cite{Debbaut2012,Schliess2014,Schwen2014,Ricken2015,White2016,Piergiovanni2017,Segovia2019}. % NOTE: Schliess2014 = VirtualLiver network (Dirk Drasdo)
Understanding fluid flow in the sinusoidal network is imperative to improve these models and make functional predictions.
Previous simulation studies of blood flow in liver microvasculature were 
either limited to large veins \cite{Schwen2014}, 
or considered only small spatial regions \cite{Debbaut2012,Piergiovanni2017}.
For the entire lobule, spanning from PV to CV, 
only continuum models had been used to simulate flow \cite{Debbaut2012,Meyer2017,Mosharaf2019}.
% NOTES on references (DONE):
%   Debbaut : small sample, not resilience; first evidence of high-current edges
%   Schwen: large veins, not resilience
%   Piergiovanni2017: small sample, not resilience
%   Mosharaf2019: porous medium theory, not resilience
%   [Schiess2014: compartment model]
Finally, the sinusoidal network can become damaged upon non-lethal toxification,
% NOTE: Hoehme2010 (=Drasdo, PNAS) considers cell death of hepatocytes, but not sinusoids
prompting the question of resilience properties of this network.
How the permeability of the sinusoidal network responds to perturbations is not known.

% SUMMARY OF RESULTS
Here, we analyze network geometry using a digital reconstruction of the sinusoidal network
based on high-resolution image data of adult mouse liver \cite{Morales2015,Morales2019}.
% The nodes of this network are well defined in terms of its branch points.
% We find that the sinusoidal network contains the minimum spanning tree of its nodes as sub-graph, 
% and is itself a subgraph of the Delaunay graph of its nodes, both to good approximation.
We develop a network generation algorithm that reproduces statistical features of the sinusoidal network
(node degree distribution, edge length distribution, mean nematic order parameter),
enabling us to simulate arbitrarily sized networks from spatially restricted biological samples
and, moreover, to explore \textit{in silico} a design space of three-dimensional networks.
While simulating random graphs with given degree distribution is a classical problem of combinatorics \cite{Bender1978},
and popular software packages exist for common models of random spatial networks \cite{Hagberg2008},
we were not aware of previous network generation algorithms that allow to prescribe both degree and edge length distribution. 

Sinusoidal networks display a weak nematic alignment along the direction of flow \cite{Debbaut2012,Morales2019,Scholich2019}.
Using our algorithm, we can systematically vary this nematic alignment in simulated networks. 
We empirically find a linear relationship between the anisotropic permeability of simulated networks
and a nematic order parameter of the networks that quantifies their anisotropic geometry.
Permeabilities allow to 
efficiently compute macroscopic, tissue-level flows using a continuum model \cite{Bonfiglio2010,Debbaut2012,Meyer2017,Mosharaf2019}, 
thus providing an effective medium theory of fluid transport.
% NOTE: Pries (e.g. 1990) modeled flow in microvasculature extensively

To quantify the fault tolerance of these networks, 
we introduce a new resilience measure, which we term \textit{permeability-at-risk} and
which quantifies changes in network permeability if a given fraction of network links is removed.
The resulting permeability-at-risk curves can be considered as a generalization of 
the bond percolation problem in the theory of random resistor networks \cite{Kirkpatrick1973,Redner2011}.
We find that simulated networks with weak nematic order
display a substantially increased permeability along the direction of nematic alignment.
If the mean nematic order parameter equals that of sinusoidal networks, 
this increased permeability does not compromise network resilience as compared to isotropic simulated networks. 
Our minimal transport model, which assumes constant and equal flow resistance per unit length for each edge, 
predicts that the distribution of computed currents is very inhomogeneous in the network,
with a few edges carrying most of the current.
This renders these networks highly vulnerable to the removal of high-current edges,
despite their resilience against random removal of edges.
In the discussion, we speculate on mechanisms such as 
shear-dependent adaptation of the diameter of sinusoids \cite{Hu2013,Chang2019,Meigel2019}, or
% NOTE: previous work on shear-dependent remodeling of arteries: Kamiya&Togawa:1980, Langille&Donnell,Science:1986(>1000 citations)
transient clogging by erythrocytes \cite{Chen2010,Savin2016}, 
which would both affect especially high-current edges, and
could homogenize the time-averaged distribution of currents in the network, 
thereby reducing the vulnerability of sinusoidal networks to the removal of high-current edges.

% ------------------------------------------------------------------------------------------------------------------
\section*{Experimental Data and Network Metrics}

\begin{figure}
\begin{center}
\includegraphics[]{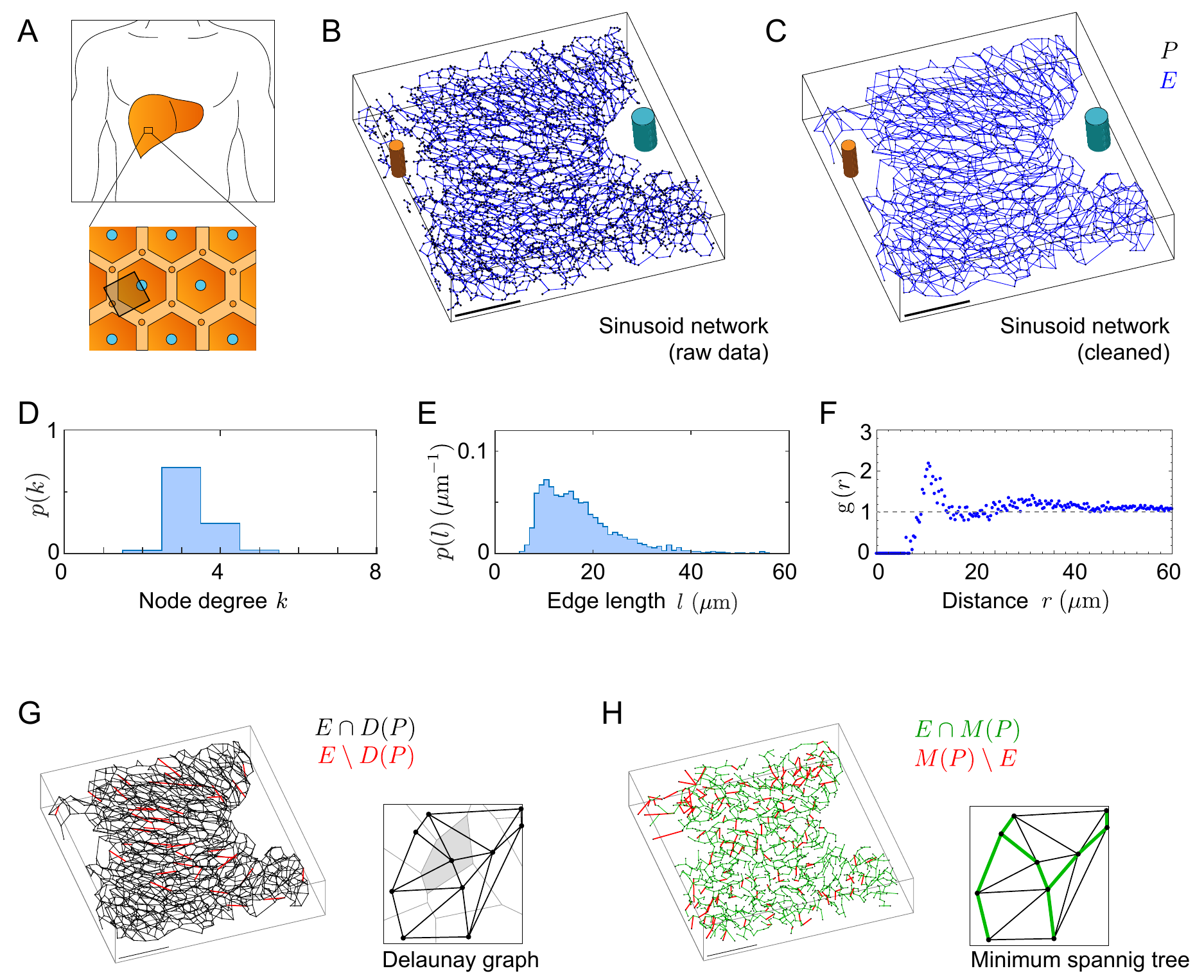}
\end{center}
\vspace{-5mm}
\caption[]{
\textbf{Sinusoidal networks in liver tissue.}
\textbf{A.} 
The liver is organized into millimeter-sized lobules, 
each containing a central vein (CV, cyan) and several portal veins (PV, orange). Blood flows from PV to CV through the plexus-like sinusoidal network.
\textbf{B.} 
Raw sinusoidal network obtained by high-resolution 3D imaging of mouse liver tissue. Shown is the network skeleton with branch points (black) and edges (blue), defining a spatial network. CV and PV shown schematically.
Scale bar: $100\,\micron$. 
% Dimensions of box (double-checked by Ben, 30.10.2018): 
% v1: $450\,\micron\,\times\,450\,\micron\,\times\,90\,\micron$
% v2,v3: $420\,\micron\,\times\,450\,\micron\,\times\,90\,\micron$ (min(pos)=[7.6500 7.8000 5.7173], max(pos)=[417.4500 439.9500 86.8122])
\textbf{C.} 
Sinusoidal network $\mathcal{N}$ obtained from the raw network by removal of degree-2 nodes and disconnected components, 
used in all subsequent analysis.
\textbf{D, E, F.} 
Distribution of node degree $p(k)$, distribution of edge lengths $p(l)$, 
and radial distribution function $g(r)$ of spatial node positions for the sinusoidal network $\mathcal{N}$, respectively.
\textbf{G.} 
The majority of edges of the sinusoidal network are contained in the Delaunay graph corresponding to its set of node positions.
Edges of the sinusoidal network not contained in the Delaunay graph are shown in red.
Inset: cartoon illustrating a Delaunay graph (black) and Voronoi tessellation (gray) of a set of points.
\textbf{H.} 
The sinusoidal network contains the majority of edges of the minimum spanning tree $M(P)$ of its set of node positions $P$.
Edges of $M(P)$ not in $E$ are shown in red.
Inset: cartoon representation of a minimum spanning tree (green).
}
\label{figure1}
\vspace{-2mm}
\end{figure}

To analyze the statistical geometry of three-dimensional microvasculature networks,
we took advantage of advances in high-resolution imaging of murine liver tissue \cite{Morales2015,Morales2019}.
Based on segmented three-dimensional image data
% (voxel size $0.3\,\micron \times 0.3\,\micron \times 0.3\,\micron$),
the skeleton of the hepatic sinusoidal network was computed using MotionTracking image analysis software \cite{Morales2015,Morales2016},
see Fig.~\ref{figure1}B.

Next, a cleaned version of the raw network data was computed:
(i) small disconnected network components not connected to the largest component were discarded,
(ii) connected nodes separated by a distance smaller than a cut-off distance $R_c=8\,\micron$
(outer diameter of sinusoids)
% optical resolution of our high-resolution imaging 0.3um [email by Hernan 12.08.2019]
were merged into a single node,
(iii) in a subsequent pruning step, dead ends were removed, leaving only nodes with node degree $d\ge 2$.
Finally, linear-chain motifs consisting of degree-two nodes in series 
were replaced by a single link with weight equal to the total length of the linear chain. 
% NOTE (discussion with Jens 09.07.2019): 
% - small number of self-loops remains after step (iv), (exp1full: n=6, exp2full: n=0, exp3full: n=1)
In rare instances, removal of a linear chain might yield triangles at the extremities of the network, 
which were also removed. 
% OLD VERSION BY JENS: In these instances the triangle is reduced to the degree three outermost degree most and the edge length is added as self-loop.
% orginal network 
%         |--------|
%   ------O--------|
% is changed to 
%         |------------------|
%   ------O--------O---------O
The remaining node points are exactly the branch points of the biological network, 
whose positions are determined with high precision.  
This clean-up procedure reduces ambiguity on small network details that were difficult to resolve
with current imaging techniques.
It provides a faithful representation of the sinusoidal network used in all subsequent analysis,
see Fig.~\ref{figure1}C.

The skeleton of the sinusoidal network is a spatial network
with set of node points $P=\{\q_i\}$ and set of edges $E$ that connect pairs of points.
We find that the sinusoidal network is a homogeneous network
with mean node degree $\langle d\rangle=3.3\pm0.6$
and a unimodal distribution $p(l)$ of edge lengths $l$ with mean $\langle l\rangle=17\pm8~\micron$,
see Fig.~\ref{figure1}CD.
% NOTE for reference: median-d = 3, median-l = 15.4 um

We characterize the relative position of node points $\q_i$ 
in terms of their normalized radial distribution function
\begin{align}
\label{eq:RDF}
g(r) = 
\frac{1}{\rho_0}\,
\left\langle
\int_{|\r-\r_0|=r}\! d^2\r\, \frac{1}{4\pi\,r^2}\,\rho(\r-\r_0)
\right\rangle_{\r_0},
\end{align}
where 
$\rho(\r)=\sum_i \delta(\r-\p_i)$ is the point density of nodes,  
$\rho_0=\langle \rho\rangle$ the mean density (with units of an inverse volume),
and $\r_0$ the position of a `central node'.
The radial distribution function is closely related to the structure factor, 
which is used in condensed matter physics to describe the packing of particles 
and which can be considered the spatial power spectral density of $\rho(\r)$ \cite{Chaikin}. % Chaikin, p. 33ff
For an ideal gas, $g(r)=1$.
Fig.~\ref{figure1}F shows $g(r)$ for the node points of the sinusoidal network.
Interestingly, we observe a peak at a characteristic distance $r\approx 10\,\micron$,
indicating a characteristic distance between the branching points of this network.
In fact, the resultant $g(r)$ resembles the radial distribution function of a fluid,
with short-range repulsion of fluid particles.
Interestingly, the observed characteristic distance of branch points corresponds to approximately half the diameter of hepatocytes \cite{Morales2019},
which is likely to set a characteristic mesh-size of the sinusoidal network.  
The ratio of branch points to number of hepatocytes is approximately $2:1$
(i.e., $n=1643$ branch points and $857$ hepatocytes for sample volume shown in Fig.~\ref{figure1}C).

% ------------------------------------------------------------------------------------------------------------------
\subsection*{The sinusoidal network is a sub-graph of its Delaunay graph}

Given the set $P$ of node positions of the sinusoidal network,
we construct the corresponding Delaunay graph with set of edges $D(P)$.
The \textit{Delaunay graph} generalizes the familiar Delaunay triangulation in the plane to three space dimensions, 
and may be considered as the graph that connects `nearest neighbors' in $P$.  
Specifically, we may assign to each node $\q\in P$ a neighborhood $V_\q$
defined as the set of all points that are closer to $\q$ than to any other point of $P\setminus\q$.
Each $V_\q$ is a polyhedron. 
Together, these polyhedra tesselate three-dimensional space, defining the so-called \textit{Voronoi tesselation} of $P$. 
Now, an edge connecting nodes $\q_1,\q_2\in P$ belongs to the Delaunay graph $D(P)$
if and only if the corresponding polyhedra $V_{\q_1}$ and $V_{\q_2}$ share a common face.
Correspondingly, $D(P)$ is the dual graph of the Voronoi tesselation.
The Delaunay graph exists and is unique provided the node points are in general position \cite{Delaunay1934}.

Remarkably, we find that the sinusoidal network is a subgraph of the Delaunay graph to very good approximation:
$99\%$ of the edges in $E$ are contained in $D(P)$, see Fig.~\ref{figure1}G.
The $1\%$ of edges contained in $E$ but not $D(P)$ (marked red) are longer than average (with a mean length of $57\pm 19\,\micron$),
yet contribute less than $10\%$ to the network permeabilities computed below.
% NOTE: 
%  Removing those edges of the sinusoidal network that do not belong to the Delaunay graph (42 out of 2697 edges for exp1)
%  reduces the Kxx permeability by 10% from Kxx=1185 mm^-2 to Kxx=1058 mm^-2 (exp1, ROI).
Thus, the sinusoidal network can be considered a network of nearest-neighbor edges,
reflecting the design requirement of a space-filling network that connects all hepatocytes in the tissue.

% ------------------------------------------------------------------------------------------------------------------
\subsection*{The sinusoidal network contains the minimum spanning tree}
The set of node points $P$ defines also a second graph, the minimum spanning tree with edges $M(P)$.
The minimum spanning tree is defined as the connected and cycle-free graph with node points $P$ for which
the sum of the edge lengths is minimal.
Note that the minimum spanning tree $M(P)$ is always a sub-graph of the Delaunay graph $D(P)$ for any set $P$ of points in Euclidean space. 
% Note proof is based on the cycle property of MST and is simple [source: wikipedia]
Remarkably, the sinusoidal network contains the minimum spanning tree of its node points as a sub-graph to good approximation:
$90\%$ of the edges $M(P)$ of the minimum spanning tree belong also to $E$, see Fig.~\ref{figure1}H.
This finding suggests a possible optimization of the sinusoid network for shortest paths.

% ------------------------------------------------------------------------------------------------------------------
\section*{A network generation algorithm for spatial networks}
\begin{figure}
\begin{center}
\includegraphics[]{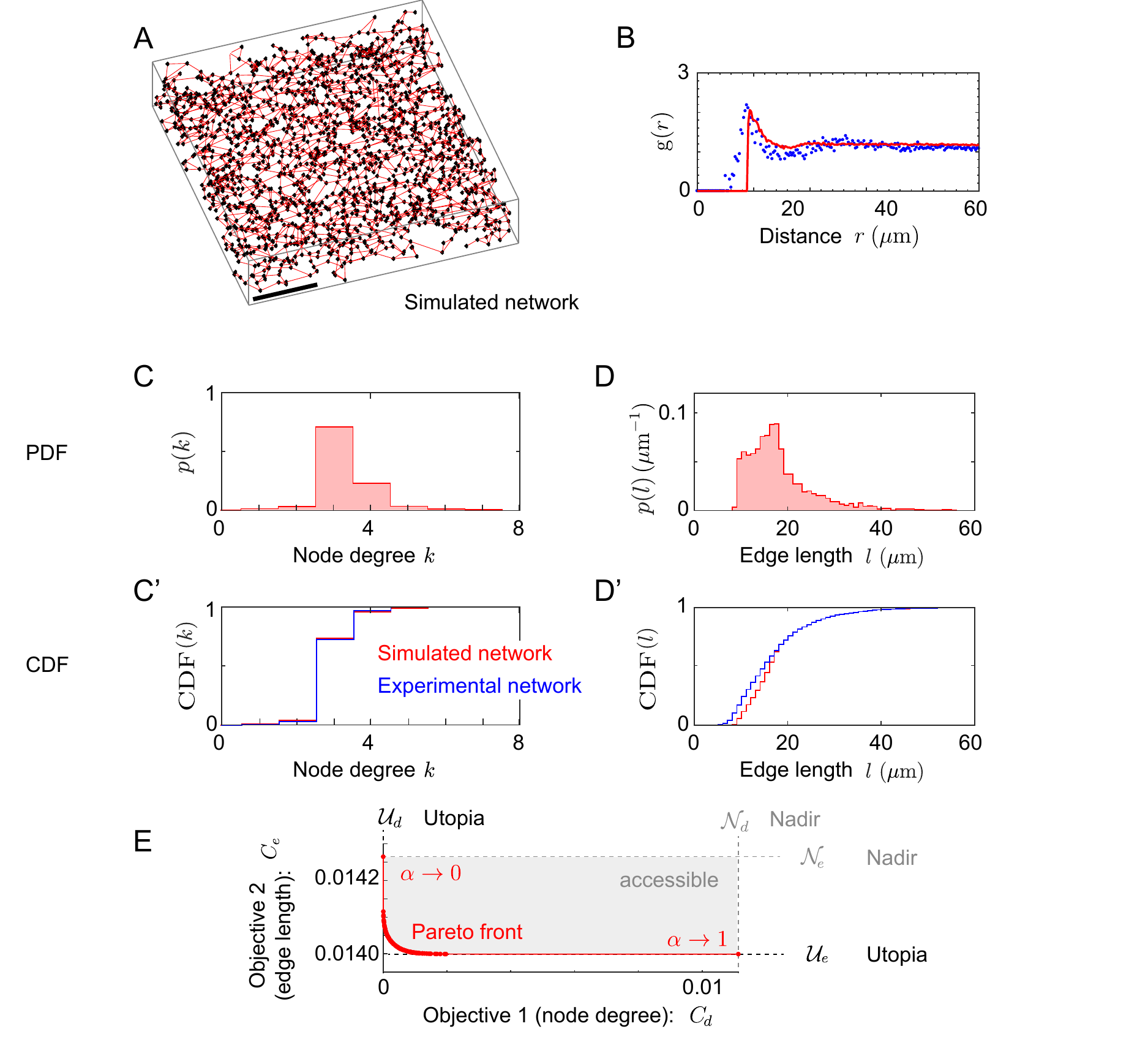}
\end{center}
\vspace{-5mm}
\caption[]{
\textbf{Simulation of networks with prescribed statistics.}
\textbf{A.}
Example of simulated network resulting from multi-objective optimization. 
The size of the simulated domain matches the dimensions of the experimental network samples
($420\,\micron \times 450\,\micron\times 90\,\micron$).
We used Monte-Carlo simulations of a packing of hard spheres to determine node positions.
\textbf{B.}
Comparison of radial distribution function $g(r)$ from sinusoidal network (black dots), and simulated node positions (red).
\textbf{C, C'.}
Probability density function (PDF) and cumulative distribution function (CDF)
for the node degree distribution $p(k)$ of a simulated network (red), and the sinusoidal network (blue).
\textbf{D, D'.} 
Same as panel C, but for the edge length distribution $p(l)$.
\textbf{E.} 
Pareto-front of multi-objective optimization.
Shown are normalized costs for each of the two objectives:
optimization of node-degree and optimization of edge-length distributions,
for varying weight $\alpha$ of the multi-objective cost function.
}
\label{figure2}
\vspace{-2mm}
\end{figure}

We developed a Monte-Carlo algorithm to generate synthetic networks that faithfully reproduce
the statistical features of spatially restricted samples of hepatic sinusoidal networks, 
using multi-objective optimization of both node degree and edge length distribution,
see Fig.~\ref{figure2}A.

Our algorithm proceeds in two steps:
first, we simulate surrogate node positions $P_\mathrm{sim}$,
followed by a second step of selecting a set of edges $E_\mathrm{sim}$ connecting these nodes.
Importantly, our analysis of sinusoidal network graphs allows us to restrict to simulated networks that
are subgraphs of the Delaunay graph $D(P_\mathrm{sim})$ and at the same time
contain the minimum spanning tree $M(P_\mathrm{sim})$,
$M(P_\mathrm{sim}) \subseteq E_\mathrm{sim} \subseteq D(P_\mathrm{sim})$.
This restriction dramatically reduces the search space
without which network optimization would be computationally unfeasible.

In the first step of the algorithm,
random node points are generated by simulating random packings of equally sized hard spheres. 
This minimal model comprises only two fit parameters 
(the volume fraction of spheres and their radius), 
and reproduces the characteristic feature of short-range repulsion
in the radial distribution function $g(r)$ for the center positions of the spheres,
see Fig.~\ref{figure2}B.
The final set of node positions $P_\mathrm{sim}$ is determined
by taking a random selection of about $20\%$ of simulated hard spheres centers,
to match the measured density $\rho_0$ of node positions in sinusoidal networks.
This random selection does not change $g(r)$,
as it scales down both denominator $\rho_0$ and numerator $\langle \int d\x\,\rho\rangle$ in Eq.~(\ref{eq:RDF}) by the same factor.
This random selection reflects volume exclusion by hepatocyte cells
(which comprise a volume fraction of about $80\%$ in liver tissue \cite{Kmiec2001}) and other cell types.

In the second step, we select a subgraph of
the Delaunay graph $D(P_\mathrm{sim})$ of the set of surrogate node positions $P_\mathrm{sim}$,
using multi-objective optimization for node degree and edge length distribution.
We introduce the cumulative probability density functions (CDF)
$\CDF(d_0)=\sum_{d\le d_0} p(d)$
and
$\CDF(l_0)=\int_0^{l_0}\! dl\, p(l)$
for node degree and edge length distribution of the sinusoidal network, respectively, 
as well as analogous CDFs $\CDF_\simu(d)$ and $\CDF_\simu(l)$ for simulated networks.
Note that $p(l)$ has dimensions of an inverse length, hence $\CDF(l)$ is dimensionless.
We define two cost functions $C_d$ and $C_e$ for the node degree and edge length distribution, respectively,
as the difference of the CDFs 
% NOTE: this is similar to the Kolmogorov-Smirnov test, which, however, considers the supremum of the differences
\begin{equation}
\label{eq:Cd}
C_d = \sum_{d=1}^\infty |\mathrm{CDF}(d)-\mathrm{CDF}_\mathrm{sim}(d)|^2
\end{equation}
and
\begin{equation}
\label{eq:Ce}
C_e = \frac{1}{R_c} \, \int_0^\infty \! dl \, |\mathrm{CDF}(l)-\mathrm{CDF}_\mathrm{sim}(l)|^2
\end{equation}
We normalized $C_e$, using the cut-off distance $R_c$ as a characteristic length scale.
% NOTE (09.07.2019): C_e-axis in Figure 2 corresponds to C_e normalized by Rc, but not normalized by utopia/nadir points
As illustration, Fig.~\ref{figure2}C,D shows probability distribution functions (PDF) and cumulative distributive functions (CDF)
for the simulated network from panel A.
The difference in CDFs provides a robust distance measure,
which is in particular robust against small shifts of the probability distributions $p(d)$ and $p(l)$. 
We also introduce a multi-objective cost function as weighted mean of the individual cost functions, 
parametrized by a weight $\alpha$
\begin{equation}
\label{eq:Calpha}
C_\alpha = \alpha C_d + (1-\alpha) C_e,
% OLD VERSION FOR REFERENCE: \mathcal{H} = \alpha \frac{J_1(\alpha) - J_1^U}{J_1^N - J_1^U}   + (1-\alpha) \frac{J_2(\alpha) - J_2^U}{J_2^N - J_2^U}
\end{equation}
% CAUTION: alpha in simulations refers to \overline{\alpha} defined below, NOT to \alpha introduced here.

We determined sets of edges $E^\ast_\simu(\alpha)$ that minimize $C_\alpha$ for given $\alpha$ by simulated annealing, see methods section for details.
Note that choosing either $\alpha=0$ or $\alpha=1$ would correspond to optimizing only a single objective, $C_e$ or $C_d$, respectively.
These single-objective optima set the \textit{utopia points}, 
i.e., the best-achievable values for each individual cost function, 
$\U_d = C_1[E^\ast_\simu(1)] = C_d[E^\ast_\simu(1)]$, and
$\U_e = C_0[E^\ast_\simu(0)] = C_l[E^\ast_\simu(0)]$, 
see Fig.~\ref{figure2}E.
The value of the respective other cost function define the \textit{nadir points} $\N_e$ and $\N_d$.
The definition of the nadir points requires to take a limit of $\alpha$,
$\alpha\nearrow 1$ or $\alpha\searrow 0$, as
$\N_e = \lim_{\alpha\nearrow 1} C_e[E^\ast_\simu(\alpha)]$, and
$\N_d = \lim_{\alpha\searrow 0} C_d[E^\ast_\simu(\alpha)]$.
Otherwise, the values of $C_e$ and $C_d$ are not well-defined for a single-objective optimization with $\alpha=1$ or $\alpha=0$, respectively. 

In the general case of multi-objective optimization with $0\le\alpha\le 1$,
the values of the individual cost functions for the optimal network are bounded
by the utopia and nadir points,
$\U_d\le C_d[E^\ast_\simu(\alpha)]\le \N_d$, and
$\U_e\le C_e[E^\ast_\simu(\alpha)]\le \N_e$.
The curve $(C_d[E^\ast_\simu(\alpha)],C_e[E^\ast_\simu(\alpha)])$, $0\le\alpha\le 1$,
defines a \textit{Pareto front} that separates a region of impossible pairs of $(C_d,C_e)$ values,
from a region of possible values.
Remarkably, this Pareto front deviates only little from the straight lines defined by $C_d=\U_d$ and $C_e=\U_e$
for intermediate values of $\alpha$.
Hence multi-objective optimization for both degree and edge length distribution
can be achieved with minimal impairment on the individual cost functions.
This shows that our algorithm is robust with respect to the particular choice of $\alpha$.

In the following, we use the value
$\alpha^\ast = (\mathcal{N}_e-\mathcal{U}_e)/(\mathcal{N}_d-\mathcal{U}_d+\mathcal{N}_e-\mathcal{U}_e)$,
which corresponds to choosing equal weights for the individual cost functions
after an appropriate normalization, see Methods section.

% FIGURE 2

% -----------------------------------------------------------------------------------------------------------------------------
\section*{Nematic order determines anisotropy of network permeability}

\begin{figure}
\begin{center}
\includegraphics[width=15cm]{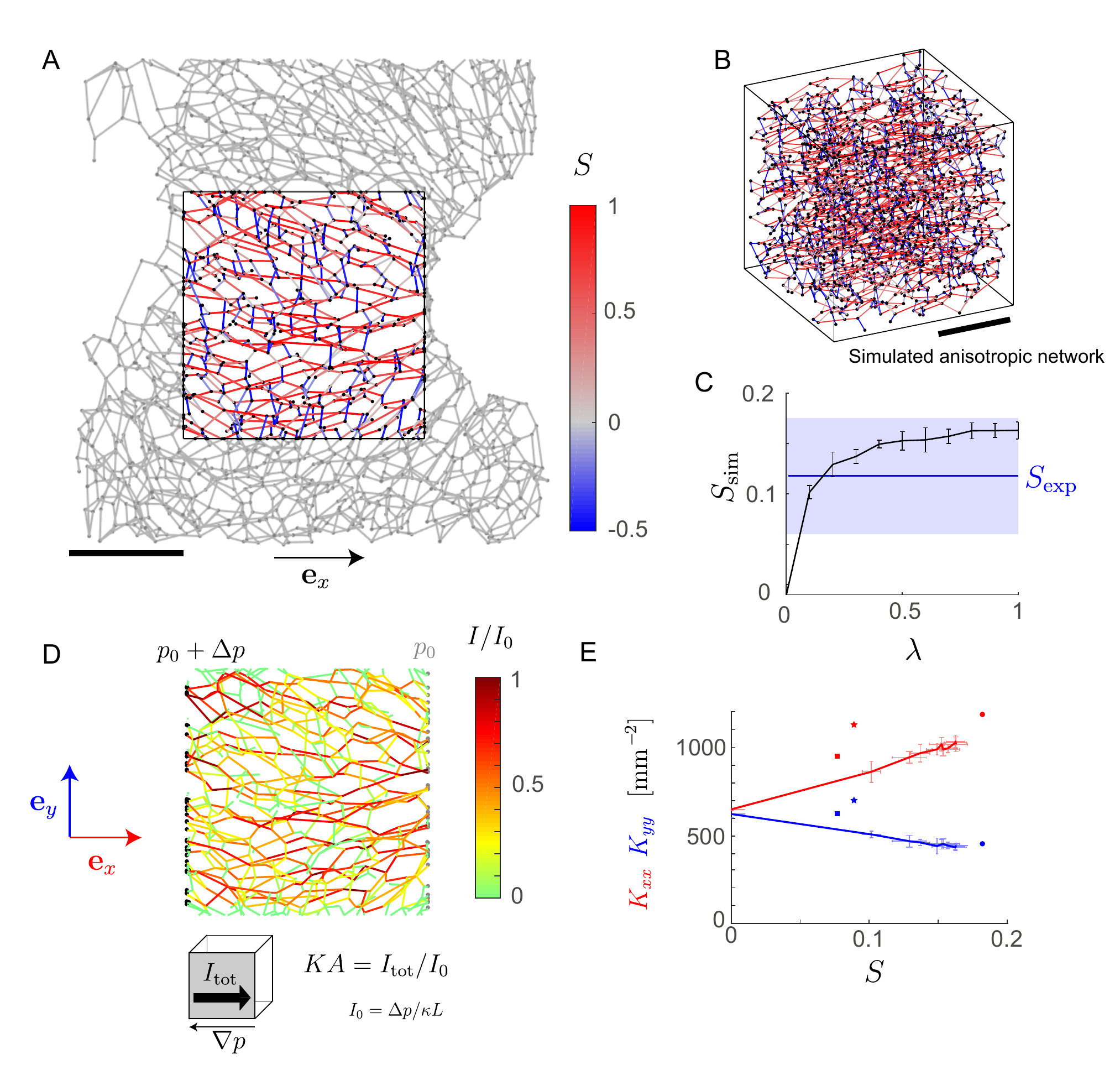}
\end{center}
\vspace{-5mm}
\caption[]{
\textbf{Nematic order determines anisotropy of network permeability.}
\textbf{A.} 
Sinusoidal network with edges inside a central region of interest color-coded according 
to their parallel alignment with the CV-PV axis $\mathbf{e}_x$
(red: parallel alignment, blue: perpendicular alignment).
Scale bar: $100\,\micron$. 
\textbf{B.} 
Simulated network with an external alignment field imposed during optimization ($\lambda=1$). 
Color-code as in panel A.
\textbf{C.} 
Nematic order parameter $S_\mathrm{sim}$ of simulated networks 
as function of external alignment field strength $\lambda$
($n=3$ realizations, mean$\pm$s.e.). 
For comparison, the nematic order parameter $S_\mathrm{exp}$ of sinusoidal networks measured along the CV-PV axis 
is shown (blue region represents mean$\pm$s.e., $n=3$).
\textbf{D.}
Computed flow through the sinusoidal network for imposed pressure difference at opposing boundary surfaces. 
\textbf{E.}
Computed permeabilities of simulated networks (solid lines)
as function of the nematic order parameter $S$
(permeability $K_{xx}$ along $\e_x$ parallel to the direction of nematic order: red, 
permeability $K_{yy}$ along $\mathbf{e}_y$ perpendicular to the direction of nematic order: blue,
mean$\pm$s.e., $n=3$ realizations). 
Symbols indicate the respective permeability sinusoidal networks 
($n=3$; filled circle correspond to network shown in panel A).
}
\label{figure3}
\vspace{-2mm}
\end{figure}

The sinusoidal network is not isotropic,
but displays weak nematic alignment along a direction parallel
to the direction of blood flow along the PV-CV axis, see Fig.~\ref{figure3}A.
Previous work has shown that the local director of sinusoidal network anisotropy
follows curvilinear flow lines, where PV and CV serve as source and sink, respectively \cite{Morales2019,Scholich2019}.
Inside a central region of the lobule, these flow lines are approximately straight,
which simplifies our subsequent analysis.
We quantify nematic alignment of the sinusoidal network inside this central region
in terms of a \textit{nematic order parameter}
\begin{equation}
\label{eq:S}
S = \left\langle \frac{3}{2} (\e_j\cdot \e_x)^2 - \frac{1}{2} \right\rangle.
\end{equation}
Here,
$\e_x$ is a unit vector pointing along the CV-PV axis
(which is parallel to the $x$ axis for the given data set), and
$\e_j$ is a unit vector parallel to the $j$-th edge.
In Eq.~(\ref{eq:S}), averaging is performed over all edges completely located inside the region of interest.
% NOTE: edges are NOT weighted by their length (discussion with Jens, 09.07.2019)
To compute $S$, 
we choose a cuboid region of interest of dimensions $L_x \times L_y \times L_z$, located in the central region of the lobule,
with cuboid edges aligned with the $x$, $y$, $z$ axes of a coordinate system, see Fig.~\ref{figure3}A.
We found $ S=0.12\pm 0.06$
(mean$\pm$s.e., $n=3$ independent data sets from different mice; $S=0.18$ for the data set in Fig.~\ref{figure3}A)
\footnote{
Note that in Eq.~(\ref{eq:S}), the contribution of each edge is independent of its length.
Since edge lengths are distributed rather homogeneously, such weighting would not change numerical results. 
However, a weighting by edge length would introduce a spurious bias towards longer edges
in simulations of anisotropic networks using Eq.~(\ref{eq:lambda}).
}.

Next, we modified our stochastic network generation algorithm,
to generate networks that likewise show nematic alignment along a specified axis.
For that aim, we compute a nematic order parameter $S_\simu$ for simulated networks
analogous to Eq.~(\ref{eq:S}) and use an augmented cost function in the optimization
\begin{equation}
\label{eq:lambda}
C_S = C_\alpha - \lambda\, \nu\, S_\simu,
\end{equation}
with interaction parameter $\lambda$. 
Here,
$\nu = \alpha^\ast (\N_d-\U_d) + (1-\alpha^\ast)(\N_e-\U_e)$
is a normalization constant introduced for numerical convenience,
corresponding to the use of a normalized cost function $\ol{C}$, see Eq.~(\ref{eq:olC}).
% NOTE: Jens used $\ol{C}_\ol{\alpha}-\lambda S$; thus, we need $\nu$ if we state analogous functional using the NON-normalized cost function
% see also Mathematica notebook 'normalization_cost.nb'
For simplicity, we do not account for a possible dependence of edge lengths on their orientation in space.
% NOTE: In fact, the empirical edge length distribution changes if we chose either edges with nematic alignment parameter S_edge<0.5 or S_edge>0.5 (mean: 16.5 um vs. 22.0 um) [Ben 14.08.2019]
Fig.~\ref{figure3}B shows an example of a simulated anisotropic network
(with nematic order parameter $S_\simu\approx 0.16$, 
similar to the value $S\approx 0.18$ measured for the experimental sinusoidal network shown in Fig.~\ref{figure3}A).
More generally, the nematic order parameter $S_\simu$ is a monotonic increasing function 
of the interaction parameter $\lambda$ in Eq.~(\ref{eq:lambda}), see Fig.~\ref{figure3}C.
The nematic order parameter $S_\simu$ of simulated networks saturates at a maximal value $S_\simu= 0.16\pm 0.01$, 
suggesting that there exists a maximal value of $S$ compatible with the given degree and edge length distribution. 

The geometric anisotropy of the sinusoidal network results in anisotropic transport properties.
We can quantify the transport properties of spatial networks 
using a minimal flow model that assumes a constant resistance $\kappa$ per unit length for each edge. 
This assumption is justified since Reynolds numbers are small \cite{Debbaut2012}.
Thus, we can model flow using Kirchhoff's laws
\begin{align}
\label{eq:Kirchhoff}
(p_{j}-p_{k}) &= \kappa\, l_{j,k}\,\, J_{j,k} \text{ for all edges connecting nodes $j$ and $k$ (Ohm's law) }, \\[1mm]
0 &= \sum_{k} J_{j,k} \text{ for all nodes $j$ (conservation of current)}.
\end{align}
% NOTE: 'kappa' not yet used by Debbaut2012, since there full 3D geometry of sinusoids was considered (yet 'k' for permeability)
Here, $J_{j,k}$ denotes the signed current through the directed edge connecting node $j$ and $k$ (with units $\mathrm{m^3\,s^{-1}}$),
$l_{j,k}$ the length of that edge, and
$\kappa$ a constant resistance per unit length (with unit $\mathrm{Pa\,m^{-4}\,s^{-1}}$),
corresponding to the simplifying assumption of a constant diameter of sinusoid tubes.

Fig.~\ref{figure3}D shows the flow computed for the chosen region of interest of the sinusoidal network.
Here, we impose a pressure difference $\Delta p$ at opposing boundaries of the region of interest as indicated.
Specifically, we identify all nodes close to the two boundary surfaces of the cuboid region
normal to the $x$ axis as either sink and source nodes, 
with $p_i=p_0$ and $p_i=p_0+\Delta p$, respectively.
Solving the linear problem defined by these boundary conditions and Eq.~(\ref{eq:Kirchhoff}),
with conservation of current at all nodes that are neither source nor sink,
yields the pressure at these remaining nodes and the currents $J_{j,k}$.
The total inflow $J_x$ at the source nodes
equals the total outflow at the sink nodes.
Contrary to intuition, we find a distribution of flows that is very inhomogeneous, see Fig.~\ref{figure3}D.
As an interesting side remark,
also an inhomogeneous distribution of flows can allow for a homogeneous supply of the tissue,
see Fig.~\ref{figureS3} for results from a minimal model of metabolic uptake from the sinusoidal network by surrounding tissue.

The \textit{permeability} of a network in $x$ direction 
is proportional to the ratio of the total current $J_x$ divided by the imposed pressure difference $\Delta p$.
We normalize the permeability by the dimension $L_x$ of the cuboid in $x$ direction, its cross-sectional area $A_x = L_y L_z$, and
the resistance per unit length $\kappa$ of individual edges,
to obtain a normalized permeability (with units of an area density) as
\begin{equation}
K_{xx} = \frac{L_x}{A_x}\,\frac{\kappa J_x}{\Delta p}.
\end{equation}
Analogously, we define $K_{yy}$ and $K_{zz}$ for the $y$ and $z$ direction, respectively.
This definition of a normalized permeability defines a purely geometrical measure of the network that is independent of $\kappa$, 
a property known as Darcy's law.
We confirmed that indeed Darcy's law holds approximately for the large samples used, 
i.e., the normalized permeability is indeed independent of the dimensions of the cuboid if boundary effects can be neglected, see Fig.~\ref{figureS1}.
For smaller sample volumes, 
computed permeabilities display a high statistical variability, but still have the same mean value.
To convey the geometric meaning of the normalized permeability, 
we consider a minimal network that consists of straight lines parallel to the $x$ axis, running from one boundary face of the cuboid to the opposing face:
there, we have $K_{xx}=K_0$, 
where $K_0$ is the area density of these lines in any cross section of the cuboid.

Fig.~\ref{figure3}E shows 
computed normalized permeabilities $K_{xx}$ and $K_{yy}$ of sinusoidal networks
in the direction of nematic alignment and normal to it, respectively.
We have
$K_{xx} = 1.1\pm 0.1\,10^3\,\mathrm{mm}^{-2}$ and % NOTE: Ben corrected unit on 03.07.2019
$K_{yy} = 0.6\pm 0.1\,10^3\,\mathrm{mm}^{-2}$ (mean$\pm$s.e., $n=3$).
The computed permeability in $z$ direction, $K_{zz} = 0.9\pm 0.3\,10^3\,\mathrm{mm}^{-2}$,
is rather unreliable due to the small dimension $L_z$ of the experimental sample in $z$ direction. 
% - adult_m1_s3_fv2_sinusoid_centrallines_simplified: Sx=0.182 Kxx=1185.2 Kyy=457.4 Kzz=556.8 (Kyy+Kzz)/2=507.1 [1/mm^2] (=circle)
% - adult_m2_s6_fv3_sinusoid_centrallines_simplified: Sx=0.077 Kxx=952.7 Kyy=626.7 Kzz=901.4 (Kyy+Kzz)/2=764.0 [1/mm^2] (=square)
% - adult_m3_s4_fv1_sinusoid_centrallines_simplified: Sx=0.089 Kxx=1129.4 Kyy=703.8 Kzz=1154.6 (Kyy+Kzz)/2=929.2 [1/mm^2] (=star)
The anisotropy ratio $K_{xx}/K_{yy}\approx 1.83$ is consistent with a previously reported value 
$2.2$ computed for a $0.15\,\micron\, \times\, 0.15\,\micron\,\times\,0.15\,\micron$ sample of the sinusoidal network imaged using $\mu$CT \cite{Debbaut2012}.

For the simulated networks, 
we find that the permeability $K_{xx}$ in the direction of nematic alignment increases linearly as a function of the nematic order parameter $S_\simu$, 
while the permeability $K_{yy}$ in normal direction decreases,
see Fig.~\ref{figure3}E.

We can estimate the resistance per unit length $\kappa$ of sinusoids 
using the Hagen-Poiseuille formula as
$\kappa = 8\mu / (\pi R^4) \approx 1.1\cdot 10^{20}\, \mathrm{Pa\,m^{-4}\,s^{-1}}$,
where
$\mu = 3.5\cdot 10^{-3}\,\mathrm{Pa\,s}$ is the dynamic viscosity of blood \cite{White2016},
$2R = 6\,\micron$ the inner diameter of sinusoids.
Given a typical pressure difference between portal and central vein
$\Delta p = 100 - 250 \,\mathrm{Pa}$ ($1-2\,\mathrm{mm\,Hg}$) \cite{Debbaut2012,Ricken2015}, 
and a typical spacing of $L=0.5\,\mathrm{mm}$, 
this value implies a volumetric flow rate of
$J=2-5\cdot 10^{-9}\,\mathrm{ml\,s^{-1}}$
through a single sinusoid connecting portal and central vein
with maximal flow velocity of 
$v=0.13-0.32\,\mathrm{mm\,s^{-1}}$, 
which validates our low Reynolds number approximation ($\mathrm{Re}\approx 10^{-4}$).
The normalized permeabilities computed here 
correspond to non-normalized permeabilities
$\mu K_{xx}/\kappa = 3.5\pm 0.3\,10{-14}\,\mathrm{m}^2$ and
$\mu K_{yy}/\kappa = 1.9\pm 0.3\,10{-14}\,\mathrm{m}^2$
with units of an area as commonly used in the theory of porous media.
These values agree within a factor of two with permeabilities along the radial and circumferential direction of the lobule
computed previously for a smaller sample volume, 
$1.56\,10{-14}\,\mathrm{m}^2$ and $1.76\,10{-14}\,\mathrm{m}^2$, respectively \cite{Debbaut2012}.
Note that the resistance $\kappa$ is very sensitive to the assumed diameter of sinusoids;
a $10\%$ decrease in $R$ would decrease the non-normalized permeability by a factor of two.

% -----------------------------------------------------------------------------------------------------------------------------
\section*{Permeability-at-risk}

\begin{figure}
\begin{center}
\includegraphics[width=15cm]{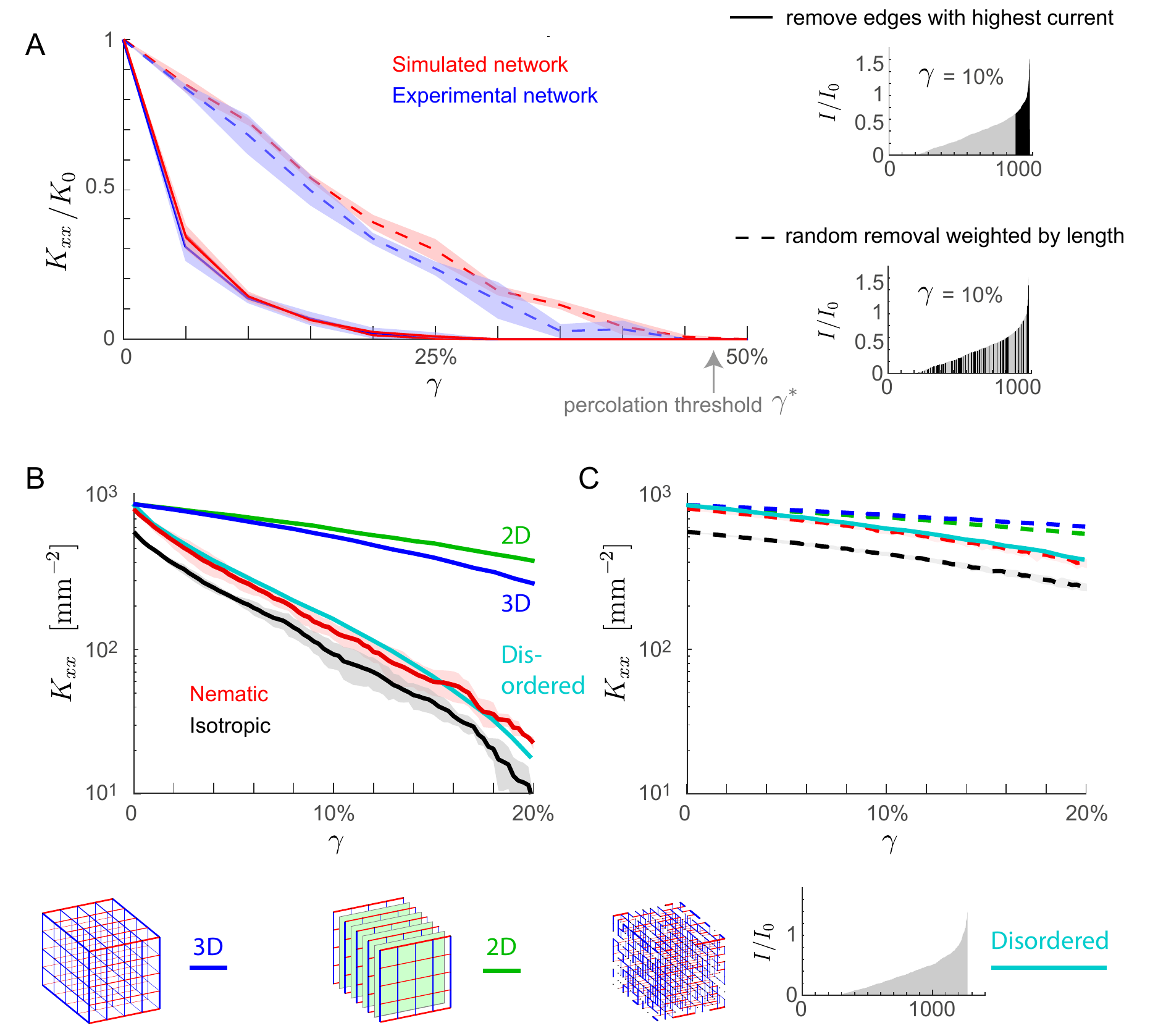}
\end{center}
\vspace{-5mm}
\caption[]{
\textbf{Resilience of simulated three-dimensional transport networks.}
% \textbf{A.} 
% We sorted network edges according to the simulated current through these edges (\textit{top}),
% and removed a fraction $\gamma=1\%$ of edges that carry the highest current in the unperturbed network (\textit{middle}). 
% The removal of edges results in a re-routing of flow (\textit{lower panel}, color-code shows change in current)
% with increase in current for some edges (red) and decrease in others (green).
% Scale bar $100\,\micron$.
\textbf{A.}
Permeability-at-risk.
Shown is the permeability of both experimental and simulated networks after perturbation
as a function of the fraction $\gamma$ of removed edges for two basic scenarios:
removal of edges that carried the highest current in the unperturbed network (solid), and
random removal of edges with removal probability proportional to their length (dashed).
Results are similar for sinusoidal networks (blue) and simulated networks (red).
Above a critical $\gamma^\ast$ (percolation threshold), the permeability drops to zero.
Permeabilities $K_{xx}$ were computed along the direction of nematic order ($x$ direction),
and normalized by the permeability $K_0$ of the unperturbed network to facilitate comparison
($K_0=1089.1\pm 121\,\mathrm{mm}^{-2}$ for sinusoidal networks, 
% exp1_ROI: 1185.2 1/mm^2
% exp2_ROI:  952.7 1/mm^2
% exp3_ROI: 1129.4 1/mm^2

$K_0=943\pm 32\,\mathrm{mm}^{-2}$ for simulated networks).
Shaded areas indicate mean$\pm$s.e. ($n=3$ networks for each condition; region of interest for sinusoidal networks as in Fig.~\ref{figure3}A).
\textbf{B.}
Logarithmic plot of permeability as function of $\gamma$ for simulated networks
for scenario \textit{i} of removal of high-current edges
for 
isotropic networks (black; $\lambda=0$), and
nematic networks (red; $\lambda=0.2$, resulting in $S\approx 0.129$, comparable to the value of sinusoidal networks). 
We compared the permeability-at-risk curve of simulated networks
to that of three simple networks:
full cubic lattice (blue; $5\times 5\times 5$, lattice spacing $31.6\,\micron$ ensuring $K_{xx}(\gamma=0)=10^3\,\mathrm{mm}^{-2}$), 
layers of square lattices (green; $5\times (5\times 5)$, lattice spacing $31.6\,\micron$), 
random sub-lattice of cubic lattice (cyan; sub-lattice of $10\times 10\times 10$, 
with filling fraction $48\%$, mean of $n=10$ realizations, lattice spacing $14.3\,\micron$).
Cartoons of the three simple networks are shown below.
\textbf{C.}
Analogous to panel B for scenario \textit{ii} of random removal of edges. 
}
\label{figure4}
\vspace{-2mm}
\end{figure}

We used the minimal flow model to study the resilience properties of sinusoidal networks, 
i.e.\ the ability of the network to support flow even after partial damage.
If edges are removed from a network, the permeability of the network decreases.%
\footnote{This known fact is easily proven by adding an edge and noting that for constant net flux $J$ 
the pressure difference $\Delta p$ must drop due to re-routing of flow in accordance to Helmholtz' theorem 
on the minimization of energy dissipation of low-Reynolds number flows \cite{HappelBrenner}.}
We quantify network resilience in terms of \textit{permeability-at-risk} curves, see Fig.~\ref{figure4}A.
Specifically, we plot the permeability of perturbed networks as a function of the fraction $\gamma$ of removed edges. 
At a critical value $\gamma^\ast$ of $\gamma$, this permeability becomes zero, indicating the \textit{percolation threshold} of the networks.
We consider two different scenarios for the removal of edges:
(i) removal of high-current edges, i.e., those edges that carried the highest current in the unperturbed network,
(ii) random removal of edges with probability proportional to their length.
We find that removing high-current edges results in a much steeper decrease of the permeability (scenario i, solid line), 
as compared to a random removal of edges (scenario ii, dashed). 
This shows that the edges that carry a high current in the unperturbed network are indeed crucial for the global transport properties of the network.
Transport by these high-current edges is only partially compensated by re-routing of flow through low-current edges.
In the absence of additional perturbations, low-current edges are dispensable 
and the network permeability decreases only little if the majority of low-current edges is removed
(e.g. if all edges with current below the $25\%$ percentile among those edges that do not touch the boundary of the region of interest are removed,
$K_{xx}$ decreases by $11.1\pm 5.1\%$).
% K_xx [1/mm^2] unperturbed, bottom25, bottom25_not_boundary (*) --> see 'exp123_K_gamma.m'
% exp1_ROI: 1185.2 1185.2  988.7 (=83.4%)
% exp2_ROI:  952.7  952.7  853.1 (=89.6%)
% exp3_ROI: 1129.4 1128.9 1057.3 (=93.6%)
However, 
a network without these low-current edges displays permeability-at-risk curves
that decay even steeper as function of $\gamma$, see Fig.~\ref{figureS2}:
low-current edges contribute at least partially to network resilience.

Remarkably, the computed permeability-at-risk curves 
match to very good extend in both scenarios for sinusoidal networks and simulated networks.
Here, the nematic order parameter $S$ of simulated networks matches the value of the experimental sinusoidal networks.

Even for a strong reduction of network permeability, 
the majority of network nodes can still be connected to both the source and the sink.
For example, we find for $\gamma=10\%$ that only $1.3\pm 2.1\%$ and $0.5\pm 0.1\%$ of nodes in the simulated network are disconnected from source or sink in scenario \textit{i} and \textit{ii}, respectively. 
As expected, these fractions increase to $100\%$ if $\gamma$ approaches the respective percolation thresholds. 
% A connected network with low permeability could still allow for transport by diffusion. 

Next, we asked how nematic order of a network influences it resilience,
using our network generation algorithm that reproduces the statistical features of sinusoidal networks, 
while allowing us to tune its nematic order parameter.
We find that nematic simulated networks (whose nematic order parameter matches those of the sinusoidal networks) 
have a higher permeability along the direction of nematic alignment than isotropic networks,
not only in the absence of perturbations, but also if edges are removed, see Fig.~\ref{figure4}BC.
We anticipate that the observed weak nematic alignment of sinusoidal networks 
represents a good compromise between increasing network permeability and decreasing network resilience.

Nonetheless, 
in scenario i, where high-current edges are removed first,
both sinusoidal networks and simulated networks display a comparatively low resilience.
To understand the origin of this vulnerability, we consider simple example networks.
In Fig.~\ref{figure4}BC, we plot resilience curves of simple regular lattices: 
a full cubic lattice (blue) and a layered stack of square lattices (green).
Both regular lattices display a much weaker reduction of network permeability upon edge removal.
We attribute this apparent resilience of regular lattices to the fact that a large subset of their edges carry the same current in the unperturbed network, 
and thus can be considered equally important. 
In contrast, disordered networks display the same vulnerability:
we tested random sub-lattices of a cubic lattice 
and observed a similar dramatic drop in network permeability if high-current edges are removed, 
see Fig.~\ref{figure4}BC (cyan).
The filling fraction of $48\%$ of this sub-lattice served as fitting parameter
that allowed to change the disorder of the network in a continuous manner.
In conclusion, a low resilience against removal of high-current edges, 
as observed in sinusoidal networks, 
appears to be a general property of disordered networks,
which are typically characterized by an inhomogeneous distribution of currents across their edges.
 
% Minimal examples: Fraction of low-current edges (criterion: |I|<0.001*I0)
%   3D: 67%, 2D: 50%, disordered 26%+/-1%
%   sim1: 1%, sim2: 1%, sim3: 1.8% (beta=0.2, alpha=0.5)
% Minimal examples: Gini coefficient
%   3D: 0.67, 2D: 0.5, disordered 0.57+/-0.01
%   sim1: 0.42, sim2: 0.42, sim3: 0.42 (beta=0.2, alpha=0.5)

% -----------------------------------------------------------------------------------------------------------------------------
\section*{Discussion and Outlook}

% SUMMARY OF RESULTS
We analyzed geometry and transport properties of microvasculatory sinusoidal networks in three-dimensional liver tissue.
% ... ALGORITHM
We drew advantage of recent advances in three-dimensional imaging and digital reconstruction of liver tissue \cite{Morales2015,Morales2019}.
These allowed us to extend previous work, which predominantly addressed two-dimensional biological transport networks
\cite{Tero2010,Alim2013,Katifori2016,Woodhouse2016,Rocks2019},
to three space dimensions.
To overcome the limitations of spatially restricted samples of sinusoidal networks, 
we developed a Monte-Carlo algorithm for multi-objective optimization to generate simulated networks 
that faithfully reproduce the statistical features of digitally reconstructed sinusoidal networks (degree and edge length distribution).
This algorithm is robust with respect to the weighting of the individual objectives.
% ... NETWORK ANISOTROPY
By varying the nematic alignment of simulated networks, 
we quantified the relationship between the nematic order parameter and the anisotropy of network permeability:
for increasing nematic order, the permeability increases along the direction of nematic alignment, 
yet decreases in the perpendicular directions, displaying a linear dependence on the nematic order parameter.
% ... PERMEABILITY AT RISK
To characterize the resilience of networks in a scale-invariant manner, we introduced the notion of \textit{permeability-at-risk}, 
i.e., we compute the residual permeability of the network as a function of the fraction $\gamma$ of edges removed.
This measure is more general 
than the familiar percolation threshold at which the permeability becomes zero almost certainly, 
or the \textit{fault tolerance} considered in Tero \textit{et al.}\cite{Tero2010}, 
which quantifies the probability of percolation 
% (i.e., permeability drops to zero),
upon removal of an individual edge.
The permeability-at-risk curves for different networks follow a characteristic, concave shape, 
monotonically decaying as function of $\gamma$ and becoming zero at the percolation threshold of the network.
Yet, individual curves do not collapse on a master curve, 
but reflect properties of individual networks.
Computed permeability-at-risk curves match for both sinusoid and simulated networks.
This serves as additional validation of our network generation algorithm.
Using simulated networks that mimic real sinusoidal networks, 
we were able to vary the nematic order parameter of the networks.
We found that weakly nematic networks not only have a higher permeability in the absence of perturbations as compared to isotropic networks, 
but also retain this property even if edges are removed. 
Thus, weakly nematic networks optimize performance without compromising resilience. 
Nonetheless, according to our minimal transport model,
sinusoidal networks and their simulated counterparts are rather vulnerable to the removal of high-current edges when compared to regular lattices.
We attribute the lower resilience of disordered lattices to the presence of a small fraction of edges that carry substantially higher currents.
Indeed, we observed the same property in random sub-lattices of regular lattices,
where likewise a small subset of edges carry high currents. 
We note that the sub-lattice represents an example of a random resistor network, 
for which a rich body of theoretical results exist \cite{Kirkpatrick1973,Redner2011}. 
% TOTAL LENGTH ?
Previous studies on optimal networks emphasized building and maintenance costs of networks,
which were modeled as functions of total network length \cite{Bottinelli2017}. 
In the case of sinusoidal networks, however, 
we anticipate that other design constraints are even more important than building costs. 
The multi-objective optimization used here to simulate synthetic networks 
adjusts both node degree and edge length distribution 
and thus controls the total length of the network 
(by fixing the total number of edges and their mean length). 

% NEWTONIAN FLUID? CLOGGING?
To compute network permeabilities,
we used a minimal transport model based on linear Kirchoff equations 
with constant resistance per unit length.
Generally, the apparent viscosity of blood flow depends on vessel diameter 
by the Fahraeus-Lindquist effect \cite{Fahraeus1931,Pries1990},
yet the variation of sinusoid diameter in the sinusoidal network is small \cite{Morales2015},
which validates our assumption of a constant viscosity.
The assumed linear relation between currents and pressure differences 
is valid for Newtonian fluids at low Reynolds numbers.
However, blood is a non-Newtonian, shear-thinning fluid, 
whose apparent viscosity decreases with flow \cite{Merrill1967,Brust2013,Lanotte2016}. 
This implies a reduced effective resistance at higher flow rates, 
which would favor an even more inhomogeneous distribution of currents within the network.
We speculate that adaptive mechanisms, 
like shear-dependent regulation of sinusoid diameter \cite{Hu2013,Chang2019,Meigel2019}, 
or sinusoid contraction dependent on local concentration of solutes \cite{Chen1997},
may facilitate a more homogeneous distribution instead.
Moreover, the diameter of red blood cells is only slightly smaller than the diameter of the sinusoids,
which can result in transient clogging of sinusoids, 
especially where flow is high \cite{Chen2010,Savin2016,Fai2017}
or where the diameter of sinusoids is reduced by adaptive mechanisms. 
% Movie S2 in the Supporting Material provides an illustratory example of this effect. 
We propose that transient clogging (corresponding to a time-varying network \cite{Karschau2018}), 
can likewise result in a more homogeneous time-averaged flow distribution.
Notwithstanding these uncertainties, our simple model serves as first approximation,
which provides a suitable tool to characterize the geometry of spatial networks, 
even if actual rates of blood flow in liver tissue should differ.

% TWO NETWORKS
In addition to the sinusoidal network, 
liver tissue comprises a second network, not considered here, the \textit{bile canaliculi network}, which transports bile fluid containing digestive enzymes \cite{Boyer2013,Dasgupta2018}
(for recent digital reconstructions see \cite{Morales2015,Meyer2017,Morales2019}). 
Like the sinusoidal network, the bile canaliculi network spans across the entire liver lobule and contacts every single hepatocyte. 
Yet, bile fluid is toxic and must not enter the blood-stream, 
hence the two networks should nowhere get too close to each other. 
In two space dimensions, these competing design requirements could only be fulfilled 
by `lines' of alternating network type, connecting source and sink.
Such an architecture would only possess low resilience to damage. 
In three space dimensions, however, parallel network layers of alternating network type are possible. 
Such a design confers high transport capacity and resilience to damage.
Indeed, signatures of such layered order were recently identified in liver tissue \cite{Morales2019}. 
It must be stressed, however, that the geometry of sinusoid and bile canaliculi networks
is much more disordered than a regular design of alternating network layers, 
as expected from self-organized networks that form in a tissue, where cells such as hepatocytes continuously divide.
Such disordered networks are characterized by an inhomogeneous distribution of currents,
where, at least in our minimal transport model, 
a small number of edges carry substantially higher currents than the majority of edges.
Low-current edges, 
even if they dispensable for the permeability of the network in the absence of perturbations
and contribute only partially to network resilience, 
play an important biological role nonetheless, e.g.\
for the uptake of metabolites from the blood-stream, which relies on diffusion through the fenestrated surface of sinusoids.
We expect that also the low-current edges contribute to the supply of hepatocytes, 
provided these are connected to high-current edges by short distances.
Future work will address the self-organization of pairs of space-filling, mutually repulsive, intertwined networks \cite{Kramer2019}, 
and study their transport and resilience properties, inspired by the bile and sinusoidal network in liver tissue.

% -----------------------------------------------------------------------------------------------------------------------------
\section*{Methods}

\subsection*{Data acquisition}
As described previously \cite{Morales2015,Morales2016,Morales2019},
fixed tissue samples of murine liver were optically cleared and treated with fluorescent antibodies for fibronectin and laminin,
thus staining the extracellular matrix surrounding the sinusoids.
Subsequently, samples were imaged at high-resolution using a multiphoton laser-scanning microscopy.
Three-dimensional image data was segmented and network skeletons computed using
MotionTracking image analysis software \cite{Morales2016}. 
The data sets analyzed in this study correspond to the same used in \cite{Morales2019}.

\subsection*{Hard sphere packing model}
We used Monte-Carlo simulations to compute the radial distribution function $g(r)=g(r;R_0,\eta)$
of a packing of equally sized hard spheres with radius $R_0$ and volume fraction $\eta$
($\eta=0.15, 0.20, 0.30, 0.35$).
In these simulations, $n=10^4$
spheres were initially positioned at regular grid positions,
then $2000$ Monte-Carlo cycles were performed, where each cycle consisted of 
testing a Monte-Carlo move with Gaussian displacement 
(standard deviation $\sigma=R_0/3$)
for each of the spheres (acceptance rate about $75\%$).

For efficient computation, we exploited the fact that the radial distribution functions scales as 
$g(\lambda r;\lambda R_0,\eta)=g(r;R_0,\eta)$
if the radius $R_0$ of the hard spheres is changed to a new value $\lambda R_0$.
Spline interpolation was used to interpolate $g(r;R_0,\eta)$ for intermediate values of $\eta$.
A fit of the radial distribution function $g(r;R_0,\eta)$ for the minimal hard sphere model
and the radial distribution function of the sinusoidal network resulted in
$R_0 = 9.0405 \micron$ and $\eta = 0.2135\, \micron^{-3}$.

In a final step, a subset of sphere positions was selected to match the node density of the sinusoidal network.
For Fig.~\ref{figure2}, 
a total of $1643$ nodes were selected in a region of dimensions 
$420\,\micron\times 450\,\micron\times 90\,\micron$, 
corresponding to a node density of $0.9659\cdot 10^5\,\mathrm{mm}^{-3}$, % by Jens, double-checked by Ben
which equals the node density of the entire sinusoidal network data set, 
including the void spaces occupied by the portal and central veins.
For Fig.~\ref{figure3}, 
a total of $1964$ nodes 
were selected in a region of dimensions 
$250\,\micron\times 250\,\micron\times 250\,\micron$, 
corresponding to a node density of $1.2571\cdot 10^5\,\mathrm{mm}^{-3}$, 
which equals the node density of the sinusoidal network
excluding the void spaces occupied by the portal and central veins.
% NOTE: 
% - This is NOT the node density inside the highlighted region of interest in Figure 3A
% - Instead, Jens used the entire volume of the cuboid, minus the volume of the large veins, which were approximated as small cuboidal regions themselves.
%   [voids occupied by large veins excluded by hand; left dx=78um, dy=151um dz=86um, right dx=100um, dy=196um]
%   NOTES by Jens: Gesamtvolumen: 1.59e7 um^3, Venenregionvolumen:  1.82e6 um^3, Gesamtzahlknoten 1643, Knotendichte: 1643/(1.307e7 um^3)  = 1.25e-4 Knoten/um^3
%
% Ben's calculation for the ROI shown in Fig. 3A
%   nNode_ROI = 681 - (43+31+63+46+26+25) = 447 [exp1, exclude nodes right on boundaries]
%   ROI: 210 x 215 x 70 [x: 100-310 um,y: 100-315 um, z: 15-85 um]
%   node density = 1.4143e-04 / um^3
%
Note that the random selection of a subset of hard sphere centers
does not change the normalized radial distribution function $g(r)$. 

For the set of simulated node positions $P_\simu$,
we computed the Delaunay graph $D(P_\simu)$ 
and the minimum spanning tree $M(P_\simu)$.
As a technical note,
the computation of the Delaunay graph $D(P_\simu)$ 
can generate artifacts at the boundaries of the simulation domain with unusually long edges. 
To avoid this, we first mirrored all nodes in the set $P_\simu$ at the $6$ planes defined by the boundaries, 
thereby obtaining an enlarged set of nodes and then
computed the Delaunay graph for this enlarged set.
Finally, we retained only edges that connected original nodes, 
while discarding those that connect to one or two mirrored nodes.

\subsection*{Multi-objective optimization}
We used simulated annealing to determine an optimal set of edges $E^\ast_\alpha$
that minimizes the multi-objective cost function $C_\alpha$ defined in Eq.~(\ref{eq:Calpha})
for a given value of $\alpha$ with $0\le\alpha\le 1$.

The optimization was initialized by a set of edges 
$E_\simu=E_0$ 
where
$E_0=M(P_\simu)\cup E_\mathrm{rand}$
comprises the minimum spanning tree $M(P_\simu)$ and
a random selection of edges $E_\mathrm{rand}\subseteq D(P_\simu)\setminus M(P_\simu)$
chosen from the set difference of the Delaunay graph and the minimum spanning tree
of the set of node positions $P_\simu$.
For the simulations shown in Fig.~\ref{figure2},
the number of edges in $E_\mathrm{rand}$ was chosen to match
the number $|E\setminus M(P)|$ of edges of the sinusoid data set
that do not belong to the minimum spanning tree.
For the simulations shown in Fig.~\ref{figure3}, 
this number $|E_\mathrm{rand}|$ was chosen 
such that volume density of edges agree for $E_0$ and $E$.
 
In each step of the optimization procedure,
we randomly selected a Monte-Carlo move that swaps a random pair of edges
$e_1\in E_\simu$ and $e_2\in D(P_\simu)\setminus E_\simu$.
This Monte-Carlo move is accepted with probability 
$\exp(-\Delta C/T)$ if $\Delta C>0$,
and with probability $1$ if $\Delta C<0$, 
resulting in an update of the set of edges $E_\simu$.
Here, $\Delta C$ is the change in the normalized cost function that would result from this move
and $T$ an effective temperature parameter.

The temperature parameter $T$ was initially set to $1$
and kept constant for an initial melting phase of $10^4$ steps.
Subsequently, $T$ was reduced by a factor of $0.998$ after every 1000 steps,
until a final temperature of $10^{-10}$ was reached.
As validation tests, we confirmed that 
(i) the cost of the final network obtained at the end the annealing procedure 
equals the cost of the minimal-cost network encountered during the entire MC-optimization,
(ii) multiple runs gave almost identical values of the minimal cost, and
(iii) the computed Pareto front is continuous and concave, as predicted by theory.
% TESTS
% - Pareto front is continuous and concave (failed in earlier versions but is ok now)
% - initial tests suggests that multiple runs give (almost) identical results
% - final state of annealing and network with minimal cost encountered during the optimization agree
To account for the difference in range of the individual cost functions, $C_d$ and $C_e$,
we used a normalized total cost function with linear rescaling $\ol{C}_\ol{\alpha}$ in the simulations, 
defined as
\begin{equation}
\label{eq:olC}
\ol{C}_\ol{\alpha} = \ol{\alpha} \ol{C}_d + (1-\ol{\alpha})\ol{C}_e.
\end{equation}
where the normalized individual cost functions read
$\ol{C}_d=(C_d-\U_d)/(\N_d-\U_d)$ and
$\ol{C}_e=(C_e-\U_e)/(\N_e-\U_e)$.
The utopia and nadir points, $\U_d$, $\U_e$, $\N_d$, $\N_e$ were determined in preliminary simulations.
The normalized cost functions satisfy 
$0\le \ol{C}_d[E^\ast_\simu(\alpha)]\le 1$ and
$0\le \ol{C}_e[E^\ast_\simu(\alpha)]\le 1$
for the set of edges $E^\ast_\simu(\alpha)$ that minimize $\ol{C}_\alpha$.
We have the linear relationship
$[\alpha(\N_d-\U_d)+(1-\alpha)(\N_e-\U_e)]\ol{C}_\ol{\alpha} + \alpha \U_d + (1-\alpha) \U_e = C_\alpha$, 
where 
$\alpha= (\N_e-\U_e)\ol{\alpha} / [ (1-\ol{\alpha})(\N_d-\U_d) + \ol{\alpha} (\N_e-\U_e) ]$.
In Fig.~\ref{figure2}E, 
we used the range $10^{-4}\le \ol{\alpha} \le 1-10^{-4}$.
The value $\ol{\alpha}=0.5$
corresponds to the value 
$\alpha^\ast = (\mathcal{N}_e-\mathcal{U}_e)/(\mathcal{N}_d-\mathcal{U}_d+\mathcal{N}_e-\mathcal{U}_e)$ % NOTE: Ben corrected formula on 04.07.2019, see Mathematica notebook 'normalization_cost.nb'
for the cost function defined in Eq.~(\ref{eq:Calpha}), 
and was used in all figures, unless stated otherwise.

% -----------------------------------------------------------------------------------------------------------------------------
\section*{Acknowledgments}

The authors are supported by the DFG through the Excellence Initiative 
by the German Federal Government and State Government 
(Clusters of Excellence \textit{cfaed} EXC 1056 and \textit{Physics of Life} EXC 2068).
J.W.\ acknowledges funding by the DAAD RISE program.
We thank 
Lutz Brusch, 
Szabolcs Horv{\'a}t, 
% Yannis Kalaidzidis, 
Felix Kramer,
Steffen Lange,
% Hernan Morales-Navarette, 
Kirstin Meyer,
Carl Modes, 
Malte Schr{\"o}der, 
Fabian Segovia-Miranda, 
Marc Timme, 
% Marino Zerial, 
as well as all members of the Biological Algorithms group for stimulating discussions.

% -----------------------------------------------------------------------------------------------------------------------------
\section*{Author contributions statement}
All authors jointly conceived the project, and played an active role in formulating the model.
J.K., A.S., J.W., B.M.F. developed the theory and performed simulations.
H. M.-N., Y.K., M.Z. provided the digital reconstruction of sinusoidal networks,
% H. M.-N., F. S.-M., Y. K., M.Z. developed the digital reconstruction of sinusoidal networks.
B.M.F.\ and J.K.\ wrote the manuscript; 
all authors commented on the manuscript draft.

\section*{Additional information}

\textbf{Supporting information:} Supporting information text and a supporting information movie accompany this paper. \newline 
\textbf{Competing interests:} The authors declare that they have no competing interests.

\bibliography{sinusoid-manuscript}

\begin{thebibliography}{10}
\expandafter\ifx\csname url\endcsname\relax
  \def\url#1{\texttt{#1}}\fi
\expandafter\ifx\csname urlprefix\endcsname\relax\def\urlprefix{URL }\fi
\expandafter\ifx\csname doiprefix\endcsname\relax\def\doiprefix{DOI }\fi
\providecommand{\bibinfo}[2]{#2}
\providecommand{\eprint}[2][]{\url{#2}}

\bibitem{Katifori2016}
\bibinfo{author}{Ronellenfitsch, H.} \& \bibinfo{author}{Katifori, E.}
\newblock \bibinfo{journal}{\bibinfo{title}{Global optimization, local
  adaptation, and the role of growth in distribution networks}}.
\newblock {\emph{\JournalTitle{Phys. Rev. Lett.}}}
  \textbf{\bibinfo{volume}{117}}, \bibinfo{pages}{138301}
  (\bibinfo{year}{2016}).

\bibitem{Tero2010}
\bibinfo{author}{Tero, A.} \emph{et~al.}
\newblock \bibinfo{journal}{\bibinfo{title}{Rules for biologically inspired
  adaptive network design}}.
\newblock {\emph{\JournalTitle{Science}}} \textbf{\bibinfo{volume}{327}},
  \bibinfo{pages}{439--442} (\bibinfo{year}{2010}).

\bibitem{Alim2013}
\bibinfo{author}{Alim, K.}, \bibinfo{author}{Amselem, G.},
  \bibinfo{author}{Peaudecerf, F.}, \bibinfo{author}{Brenner, M.~P.} \&
  \bibinfo{author}{Pringle, A.}
\newblock \bibinfo{journal}{\bibinfo{title}{Random network peristalsis in
  physarum polycephalum organizes fluid flows across an individual}}.
\newblock {\emph{\JournalTitle{Proceedings of the National Academy of
  Sciences}}} \textbf{\bibinfo{volume}{110}}, \bibinfo{pages}{13306--13311}
  (\bibinfo{year}{2013}).

\bibitem{Perna2014}
\bibinfo{author}{Perna, A.} \& \bibinfo{author}{Latty, T.}
\newblock \bibinfo{journal}{\bibinfo{title}{Animal transportation networks}}.
\newblock {\emph{\JournalTitle{J. Roy. Soc. Interface}}}
  \textbf{\bibinfo{volume}{11}}, \bibinfo{pages}{20140334}
  (\bibinfo{year}{2014}).

\bibitem{Seybold2007}
\bibinfo{author}{Seybold, H.}, \bibinfo{author}{Andrade, J.~S.} \&
  \bibinfo{author}{Herrmann, H.~J.}
\newblock \bibinfo{journal}{\bibinfo{title}{Modeling river delta formation}}.
\newblock {\emph{\JournalTitle{Proc. Natl. Acad. Sci. U.S.A.}}}
  \textbf{\bibinfo{volume}{104}}, \bibinfo{pages}{16804--16809}
  (\bibinfo{year}{2007}).

\bibitem{Radjai1996}
\bibinfo{author}{Radjai, F.}, \bibinfo{author}{Jean, M.},
  \bibinfo{author}{Moreau, J.-J.} \& \bibinfo{author}{Roux, S.}
\newblock \bibinfo{journal}{\bibinfo{title}{Force distributions in dense
  two-dimensional granular systems}}.
\newblock {\emph{\JournalTitle{Phys. Rev. Lett.}}}
  \textbf{\bibinfo{volume}{77}}, \bibinfo{pages}{274} (\bibinfo{year}{1996}).

\bibitem{Kmiec2001}
\bibinfo{author}{Kmie{\'c}, Z.}
\newblock \bibinfo{title}{Introduction: Morphology of the liver lobule}.
\newblock In \emph{\bibinfo{booktitle}{Cooperation of liver cells in health and
  disease}}, \bibinfo{pages}{1--6} (\bibinfo{publisher}{Springer},
  \bibinfo{year}{2001}).

\bibitem{Katifori2010}
\bibinfo{author}{Katifori, E.},
  \bibinfo{author}{Sz\"oll\ifmmode~\mbox{\H{o}}\else \H{o}\fi{}si, G.~J.} \&
  \bibinfo{author}{Magnasco, M.~O.}
\newblock \bibinfo{journal}{\bibinfo{title}{Damage and fluctuations induce
  loops in optimal transport networks}}.
\newblock {\emph{\JournalTitle{Phys. Rev. Lett.}}}
  \textbf{\bibinfo{volume}{104}}, \bibinfo{pages}{048704}
  (\bibinfo{year}{2010}).

\bibitem{Bottinelli2017}
\bibinfo{author}{Bottinelli, A.}, \bibinfo{author}{Louf, R.} \&
  \bibinfo{author}{Gherardi, M.}
\newblock \bibinfo{journal}{\bibinfo{title}{Balancing building and maintenance
  costs in growing transport networks}}.
\newblock {\emph{\JournalTitle{Physical Review E}}}
  \textbf{\bibinfo{volume}{96}}, \bibinfo{pages}{032316}
  (\bibinfo{year}{2017}).

\bibitem{Farr2014}
\bibinfo{author}{Farr, R.~S.}, \bibinfo{author}{Harer, J.~L.} \&
  \bibinfo{author}{Fink, T.~M.}
\newblock \bibinfo{journal}{\bibinfo{title}{Easily repairable networks:
  Reconnecting nodes after damage}}.
\newblock {\emph{\JournalTitle{Phys. Rev. Lett.}}}
  \textbf{\bibinfo{volume}{113}}, \bibinfo{pages}{138701}
  (\bibinfo{year}{2014}).

\bibitem{Modes2016}
\bibinfo{author}{Modes, C.~D.}, \bibinfo{author}{Magnasco, M.~O.} \&
  \bibinfo{author}{Katifori, E.}
\newblock \bibinfo{journal}{\bibinfo{title}{Extracting hidden hierarchies in 3d
  distribution networks}}.
\newblock {\emph{\JournalTitle{Phys. Rev. X}}} \textbf{\bibinfo{volume}{6}},
  \bibinfo{pages}{031009} (\bibinfo{year}{2016}).

\bibitem{Morales2015}
\bibinfo{author}{Morales-Navarrete, H.} \emph{et~al.}
\newblock \bibinfo{journal}{\bibinfo{title}{A versatile pipeline for the
  multi-scale digital reconstruction and quantitative analysis of 3d tissue
  architecture}}.
\newblock {\emph{\JournalTitle{eLife}}} \textbf{\bibinfo{volume}{4}},
  \bibinfo{pages}{e11214} (\bibinfo{year}{2015}).

\bibitem{Morales2019}
\bibinfo{author}{Morales-Navarrete, H.} \emph{et~al.}
\newblock \bibinfo{journal}{\bibinfo{title}{Liquid-crystal organization of
  liver tissue}}.
\newblock {\emph{\JournalTitle{eLife}}} \textbf{\bibinfo{volume}{8}},
  \bibinfo{pages}{e44860} (\bibinfo{year}{2019}).

\bibitem{Debbaut2012}
\bibinfo{author}{Debbaut, C.} \emph{et~al.}
\newblock \bibinfo{journal}{\bibinfo{title}{Perfusion characteristics of the
  human hepatic microcirculation based on three-dimensional reconstructions and
  computational fluid dynamic analysis}}.
\newblock {\emph{\JournalTitle{J. Biomech. Engin.}}}
  \textbf{\bibinfo{volume}{134}}, \bibinfo{pages}{011003}
  (\bibinfo{year}{2012}).

\bibitem{Schliess2014}
\bibinfo{author}{Schliess, F.} \emph{et~al.}
\newblock \bibinfo{journal}{\bibinfo{title}{Integrated metabolic
  spatial-temporal model for the prediction of ammonia detoxification during
  liver damage and regeneration}}.
\newblock {\emph{\JournalTitle{Hepatology}}} \textbf{\bibinfo{volume}{60}},
  \bibinfo{pages}{2040--2051} (\bibinfo{year}{2014}).

\bibitem{Schwen2014}
\bibinfo{author}{Schwen, L.~O.} \emph{et~al.}
\newblock \bibinfo{journal}{\bibinfo{title}{Spatio-temporal simulation of first
  pass drug perfusion in the liver}}.
\newblock {\emph{\JournalTitle{PLoS Comp. Biol.}}}
  \textbf{\bibinfo{volume}{10}}, \bibinfo{pages}{e1003499}
  (\bibinfo{year}{2014}).

\bibitem{Ricken2015}
\bibinfo{author}{Ricken, T.} \emph{et~al.}
\newblock \bibinfo{journal}{\bibinfo{title}{Modeling function--perfusion
  behavior in liver lobules including tissue, blood, glucose, lactate and
  glycogen by use of a coupled two-scale pde--ode approach}}.
\newblock {\emph{\JournalTitle{Biomechanics and modeling in mechanobiology}}}
  \textbf{\bibinfo{volume}{14}}, \bibinfo{pages}{515--536}
  (\bibinfo{year}{2015}).

\bibitem{White2016}
\bibinfo{author}{White, D.}, \bibinfo{author}{Coombe, D.},
  \bibinfo{author}{Rezania, V.} \& \bibinfo{author}{Tuszynski, J.}
\newblock \bibinfo{journal}{\bibinfo{title}{Building a 3d virtual liver:
  Methods for simulating blood flow and hepatic clearance on 3d structures}}.
\newblock {\emph{\JournalTitle{PloS one}}} \textbf{\bibinfo{volume}{11}},
  \bibinfo{pages}{e0162215} (\bibinfo{year}{2016}).

\bibitem{Piergiovanni2017}
\bibinfo{author}{Piergiovanni, M.} \emph{et~al.}
\newblock \bibinfo{journal}{\bibinfo{title}{Microcirculation in the murine
  liver: a computational fluid dynamic model based on 3d reconstruction from in
  vivo microscopy}}.
\newblock {\emph{\JournalTitle{J. Biomech.}}} \textbf{\bibinfo{volume}{63}},
  \bibinfo{pages}{125--134} (\bibinfo{year}{2017}).

\bibitem{Segovia2019}
\bibinfo{author}{Segovia-Miranda, F.} \emph{et~al.}
\newblock \bibinfo{journal}{\bibinfo{title}{Three-dimensional spatially
  resolved geometrical and functional models of human liver tissue reveal new
  aspects of nafld progression}}.
\newblock {\emph{\JournalTitle{Nature Medicine}}} \bibinfo{pages}{1--9}
  (\bibinfo{year}{2019}).

\bibitem{Meyer2017}
\bibinfo{author}{Meyer, K.} \emph{et~al.}
\newblock \bibinfo{journal}{\bibinfo{title}{A predictive 3d multi-scale model
  of biliary fluid dynamics in the liver lobule}}.
\newblock {\emph{\JournalTitle{Cell Systems}}} \textbf{\bibinfo{volume}{4}},
  \bibinfo{pages}{277--290} (\bibinfo{year}{2017}).

\bibitem{Mosharaf2019}
\bibinfo{author}{Mosharaf-Dehkordi, M.}
\newblock \bibinfo{journal}{\bibinfo{title}{A fully coupled porous media and
  channels flow approach for simulation of blood and bile flow through the
  liver lobules}}.
\newblock {\emph{\JournalTitle{Comp. Meth. Biomech. Biomed. Engin.}}}
  \textbf{\bibinfo{volume}{22}}, \bibinfo{pages}{901--915}
  (\bibinfo{year}{2019}).

\bibitem{Bender1978}
\bibinfo{author}{Bender, E.~A.} \& \bibinfo{author}{Canfield, E.~R.}
\newblock \bibinfo{journal}{\bibinfo{title}{The asymptotic number of labeled
  graphs with given degree sequences}}.
\newblock {\emph{\JournalTitle{J. Combinat. Theory A}}}
  \textbf{\bibinfo{volume}{24}}, \bibinfo{pages}{296--307}
  (\bibinfo{year}{1978}).

\bibitem{Hagberg2008}
\bibinfo{author}{Hagberg, A.}, \bibinfo{author}{Swart, P.} \&
  \bibinfo{author}{S~Chult, D.}
\newblock \bibinfo{title}{Exploring network structure, dynamics, and function
  using networkx}.
\newblock \bibinfo{type}{Tech. Rep.}, \bibinfo{institution}{Los Alamos National
  Lab.(LANL), Los Alamos, NM (United States)} (\bibinfo{year}{2008}).

\bibitem{Scholich2019}
\bibinfo{author}{Scholich, A.} \emph{et~al.}
\newblock \bibinfo{title}{Quantification of nematic cell polarity in
  three-dimensional tissues} (\bibinfo{year}{2019}).
\newblock \bibinfo{note}{Preprint arXiv:1904.08886; available online at
  \url{https://arxiv.org/abs/1904.08886}}.

\bibitem{Bonfiglio2010}
\bibinfo{author}{Bonfiglio, A.}, \bibinfo{author}{Leungchavaphongse, K.},
  \bibinfo{author}{Repetto, R.} \& \bibinfo{author}{Siggers, J.~H.}
\newblock \bibinfo{journal}{\bibinfo{title}{{Mathematical Modeling of the
  Circulation in the Liver Lobule}}}.
\newblock {\emph{\JournalTitle{J. Biomech. Engin.}}}
  \textbf{\bibinfo{volume}{132}} (\bibinfo{year}{2010}).

\bibitem{Kirkpatrick1973}
\bibinfo{author}{Kirkpatrick, S.}
\newblock \bibinfo{journal}{\bibinfo{title}{Percolation and conduction}}.
\newblock {\emph{\JournalTitle{Rev. Mod. Phys.}}}
  \textbf{\bibinfo{volume}{45}}, \bibinfo{pages}{574} (\bibinfo{year}{1973}).

\bibitem{Redner2011}
\bibinfo{author}{Redner, S.}
\newblock \emph{\bibinfo{title}{Fractal and Multifractal Scaling of Electrical
  Conduction in Random Resistor Networks}}, \bibinfo{pages}{446--462}
  (\bibinfo{publisher}{Springer New York}, \bibinfo{address}{New York, NY},
  \bibinfo{year}{2011}).

\bibitem{Hu2013}
\bibinfo{author}{Hu, D.} \& \bibinfo{author}{Cai, D.}
\newblock \bibinfo{journal}{\bibinfo{title}{Adaptation and optimization of
  biological transport networks}}.
\newblock {\emph{\JournalTitle{Phys. Rev. Lett.}}}
  \textbf{\bibinfo{volume}{111}}, \bibinfo{pages}{138701}
  (\bibinfo{year}{2013}).

\bibitem{Chang2019}
\bibinfo{author}{Chang, S.-S.} \& \bibinfo{author}{Roper, M.}
\newblock \bibinfo{journal}{\bibinfo{title}{Microvscular networks with uniform
  flow}}.
\newblock {\emph{\JournalTitle{J. Theoret. Biol.}}}
  \textbf{\bibinfo{volume}{462}}, \bibinfo{pages}{48--64}
  (\bibinfo{year}{2019}).

\bibitem{Meigel2019}
\bibinfo{author}{Meigel, F.~J.}, \bibinfo{author}{Cha, P.},
  \bibinfo{author}{Brenner, M.~P.} \& \bibinfo{author}{Alim, K.}
\newblock \bibinfo{journal}{\bibinfo{title}{Robust increase in supply by vessel
  dilation in globally coupled microvasculature}}.
\newblock {\emph{\JournalTitle{Phys. Rev. Lett.}}}
  \textbf{\bibinfo{volume}{123}} (\bibinfo{year}{2019}).

\bibitem{Chen2010}
\bibinfo{author}{Chen, Y.-C.}, \bibinfo{author}{Chen, G.-Y.},
  \bibinfo{author}{Lin, Y.-C.} \& \bibinfo{author}{Wang, G.-J.}
\newblock \bibinfo{journal}{\bibinfo{title}{A lab-on-a-chip capillary network
  for red blood cell hydrodynamics}}.
\newblock {\emph{\JournalTitle{Microfluid. \& Nanofluid.}}}
  \textbf{\bibinfo{volume}{9}}, \bibinfo{pages}{585--591}
  (\bibinfo{year}{2010}).

\bibitem{Savin2016}
\bibinfo{author}{Savin, T.}, \bibinfo{author}{Bandi, M.} \&
  \bibinfo{author}{Mahadevan, L.}
\newblock \bibinfo{journal}{\bibinfo{title}{Pressure-driven occlusive flow of a
  confined red blood cell}}.
\newblock {\emph{\JournalTitle{Soft matter}}} \textbf{\bibinfo{volume}{12}},
  \bibinfo{pages}{562--573} (\bibinfo{year}{2016}).

\bibitem{Morales2016}
\bibinfo{author}{Morales-Navarrete, H.}, \bibinfo{author}{Nonaka, H.},
  \bibinfo{author}{Segovia-Miranda, F.}, \bibinfo{author}{Zerial, M.} \&
  \bibinfo{author}{Kalaidzidis, Y.}
\newblock \bibinfo{title}{Automatic recognition and characterization of
  different non-parenchymal cells in liver tissue}.
\newblock In \emph{\bibinfo{booktitle}{2016 IEEE 13th International Symposium
  on Biomedical Imaging (ISBI)}}, \bibinfo{pages}{536--540}
  (\bibinfo{organization}{IEEE}, \bibinfo{year}{2016}).

\bibitem{Chaikin}
\bibinfo{author}{Chaikin, P.~M.}, \bibinfo{author}{Lubensky, T.~C.} \&
  \bibinfo{author}{Witten, T.~A.}
\newblock \emph{\bibinfo{title}{Principles of {C}ondensed {M}atter {P}hysics}},
  vol.~\bibinfo{volume}{1} (\bibinfo{publisher}{Cambridge University Press
  Cambridge}, \bibinfo{year}{1995}).

\bibitem{Delaunay1934}
\bibinfo{author}{Delaunay, B.} \emph{et~al.}
\newblock \bibinfo{journal}{\bibinfo{title}{Sur la sphere vide}}.
\newblock {\emph{\JournalTitle{Izv. Akad. Nauk SSSR, Otdelenie Matematicheskii
  i Estestvennyka Nauk}}} \textbf{\bibinfo{volume}{7}}, \bibinfo{pages}{1--2}
  (\bibinfo{year}{1934}).

\bibitem{HappelBrenner}
\bibinfo{author}{Happel, J.} \& \bibinfo{author}{Brenner, H.}
\newblock \emph{\bibinfo{title}{Low Reynolds number hydrodynamics: with special
  applications to particulate media}}, vol.~\bibinfo{volume}{1}
  (\bibinfo{publisher}{Springer}, \bibinfo{year}{2012}).

\bibitem{Woodhouse2016}
\bibinfo{author}{Woodhouse, F.~G.}, \bibinfo{author}{Forrow, A.},
  \bibinfo{author}{Fawcett, J.~B.} \& \bibinfo{author}{Dunkel, J.}
\newblock \bibinfo{journal}{\bibinfo{title}{Stochastic cycle selection in
  active flow networks}}.
\newblock {\emph{\JournalTitle{Proc. Natl. Acad. Sci. U.S.A.}}}
  \textbf{\bibinfo{volume}{113}}, \bibinfo{pages}{8200--8205}
  (\bibinfo{year}{2016}).

\bibitem{Rocks2019}
\bibinfo{author}{Rocks, J.~W.}, \bibinfo{author}{Ronellenfitsch, H.},
  \bibinfo{author}{Liu, A.~J.}, \bibinfo{author}{Nagel, S.~R.} \&
  \bibinfo{author}{Katifori, E.}
\newblock \bibinfo{journal}{\bibinfo{title}{Limits of multifunctionality in
  tunable networks}}.
\newblock {\emph{\JournalTitle{Proc. Natl. Acad. Sci. U.S.A.}}}
  \textbf{\bibinfo{volume}{116}}, \bibinfo{pages}{2506--2511}
  (\bibinfo{year}{2019}).

\bibitem{Fahraeus1931}
\bibinfo{author}{Fahraeus, R.} \& \bibinfo{author}{Lindqvist, T.}
\newblock \bibinfo{journal}{\bibinfo{title}{The viscosity of the blood in
  narrow capillary tubes}}.
\newblock {\emph{\JournalTitle{Am. J. Physiol.}}}
  \textbf{\bibinfo{volume}{96}}, \bibinfo{pages}{562--568}
  (\bibinfo{year}{1931}).

\bibitem{Pries1990}
\bibinfo{author}{Pries, A.~R.}, \bibinfo{author}{Secomb, T.~W.},
  \bibinfo{author}{Gaehtgens, P.} \& \bibinfo{author}{Gross, J.~F.}
\newblock \bibinfo{journal}{\bibinfo{title}{Blood flow in microvascular
  networks. experiments and simulation.}}
\newblock {\emph{\JournalTitle{Circul. Res.}}} \textbf{\bibinfo{volume}{67}},
  \bibinfo{pages}{826--834} (\bibinfo{year}{1990}).

\bibitem{Merrill1967}
\bibinfo{author}{Merrill, E.~W.} \& \bibinfo{author}{Pelletier, G.~A.}
\newblock \bibinfo{journal}{\bibinfo{title}{Viscosity of human blood:
  transition from {N}ewtonian to non-{N}ewtonian.}}
\newblock {\emph{\JournalTitle{J. Appl. Physiol.}}}
  \textbf{\bibinfo{volume}{23}}, \bibinfo{pages}{178--182}
  (\bibinfo{year}{1967}).

\bibitem{Brust2013}
\bibinfo{author}{Brust, M.} \emph{et~al.}
\newblock \bibinfo{journal}{\bibinfo{title}{Rheology of human blood plasma:
  Viscoelastic versus newtonian behavior}}.
\newblock {\emph{\JournalTitle{Phys. Rev. Lett.}}}
  \textbf{\bibinfo{volume}{110}}, \bibinfo{pages}{078305}
  (\bibinfo{year}{2013}).

\bibitem{Lanotte2016}
\bibinfo{author}{Lanotte, L.} \emph{et~al.}
\newblock \bibinfo{journal}{\bibinfo{title}{Red cells' dynamic morphologies
  govern blood shear thinning under microcirculatory flow conditions}}.
\newblock {\emph{\JournalTitle{Proc. Natl. Acad. Sci. U.S.A.}}}
  \textbf{\bibinfo{volume}{113}}, \bibinfo{pages}{13289--13294}
  (\bibinfo{year}{2016}).

\bibitem{Chen1997}
\bibinfo{author}{Chen, Q.} \& \bibinfo{author}{Anderson, D.~R.}
\newblock \bibinfo{journal}{\bibinfo{title}{Effect of co2 on intracellular ph
  and contraction of retinal capillary pericytes.}}
\newblock {\emph{\JournalTitle{Investigative Ophthalmology \& Visual Science}}}
  \textbf{\bibinfo{volume}{38}}, \bibinfo{pages}{643--651}
  (\bibinfo{year}{1997}).

\bibitem{Fai2017}
\bibinfo{author}{Fai, T.~G.}, \bibinfo{author}{Leo-Macias, A.},
  \bibinfo{author}{Stokes, D.~L.} \& \bibinfo{author}{Peskin, C.~S.}
\newblock \bibinfo{journal}{\bibinfo{title}{Image-based model of the spectrin
  cytoskeleton for red blood cell simulation}}.
\newblock {\emph{\JournalTitle{PLoS Comp. Biol.}}}
  \textbf{\bibinfo{volume}{13}}, \bibinfo{pages}{e1005790}
  (\bibinfo{year}{2017}).

\bibitem{Karschau2018}
\bibinfo{author}{Karschau, J.}, \bibinfo{author}{Zimmerling, M.} \&
  \bibinfo{author}{Friedrich, B.~M.}
\newblock \bibinfo{journal}{\bibinfo{title}{Renormalization group theory for
  percolation in time-varying networks}}.
\newblock {\emph{\JournalTitle{Sci. Rep.}}} \textbf{\bibinfo{volume}{8}},
  \bibinfo{pages}{8011} (\bibinfo{year}{2018}).

\bibitem{Boyer2013}
\bibinfo{author}{Boyer, J.~L.}
\newblock \bibinfo{journal}{\bibinfo{title}{Bile formation and secretion}}.
\newblock {\emph{\JournalTitle{Comprehensive Physiology}}}
  \textbf{\bibinfo{volume}{3}}, \bibinfo{pages}{1035--1078}
  (\bibinfo{year}{2013}).

\bibitem{Dasgupta2018}
\bibinfo{author}{Dasgupta, S.}, \bibinfo{author}{Gupta, K.},
  \bibinfo{author}{Zhang, Y.}, \bibinfo{author}{Viasnoff, V.} \&
  \bibinfo{author}{Prost, J.}
\newblock \bibinfo{journal}{\bibinfo{title}{Physics of lumen growth}}.
\newblock {\emph{\JournalTitle{Proc. Natl. Acad. Sci. U.S.A.}}}
  \textbf{\bibinfo{volume}{115}}, \bibinfo{pages}{E4751--E4757}
  (\bibinfo{year}{2018}).

\bibitem{Kramer2019}
\bibinfo{author}{Kramer, F.} \& \bibinfo{author}{Modes, C.}
\newblock \bibinfo{title}{How to pare a pair: Topology control and pruning in
  intertwined complex networks} (\bibinfo{year}{2019}).
\newblock \bibinfo{note}{Available online at
  \url{https://www.biorxiv.org/content/10.1101/763649v1}}.

\bibitem{Meigel2018}
\bibinfo{author}{Meigel, F.~J.} \& \bibinfo{author}{Alim, K.}
\newblock \bibinfo{journal}{\bibinfo{title}{Flow rate of transport network
  controls uniform metabolite supply to tissue}}.
\newblock {\emph{\JournalTitle{J Roy. Soc. Int.}}}
  \textbf{\bibinfo{volume}{15}}, \bibinfo{pages}{20180075}
  (\bibinfo{year}{2018}).

\end{thebibliography}

\newpage

\vspace{15mm}
\subsection*{Supporting Information}

\renewcommand{\theequation}{S\arabic{equation}}    
\setcounter{equation}{0}  % reset counter     
\renewcommand{\thefigure}{S\arabic{figure}}    
\setcounter{figure}{0}  % reset counter     

Supporting information for Karschau et al.: 
\textit{
Resilience of three-dimensional sinusoidal networks in liver tissue.
}

\vspace{3mm}

\textit{Supporting information movie S1.}
Sinusoidal network from Fig.~\ref{figure1}C, corresponding to part of a liver lobule spanning from central to portal vein.
The centerlines of the sinusoidal network are obtained from high-resolution three-dimensional imaging of mouse liver tissue 
and digital reconstruction \cite{Morales2015,Morales2019}.

% \textit{Supporting material movie S2.}
% Fluorescence microscope recording of blood flow through murine sinusoidal network,
% with fluorescent marker dextran injected into the blood stream of the animals.
% Red channel: dextran, which marks the marks liquid fraction of blood;
% green channel: CFDA, which transiently marks bile canaliculi;
% blue channel: Hoechst staining, which marks cell nuclei (e.g., hepatocytes) in the live animal.
% The diameter of cell nuclei is about $10\,\micron$.
% White and red blood cells are visible in the dextran-labeled blood stream as black areas
% (round black areas: mostly white blood cells, disk-like black areas perpendicular to the flow: red blood cells).
% In several instances, 
% a temporary slowdown of cell movement (transient clogging) with subsequent acceleration
% is visible, which was a typical phenomenon in several recordings.
% Experimental methods are described in detail in Ref. \cite{Meyer2017},
% where additional movies are shown.
% We are grateful to Kirstin Meyer for providing the raw data for movie S2. [TODO: move to Acknowledgements.]

\begin{figure}[hb]
\begin{center}
\includegraphics[width=15cm]{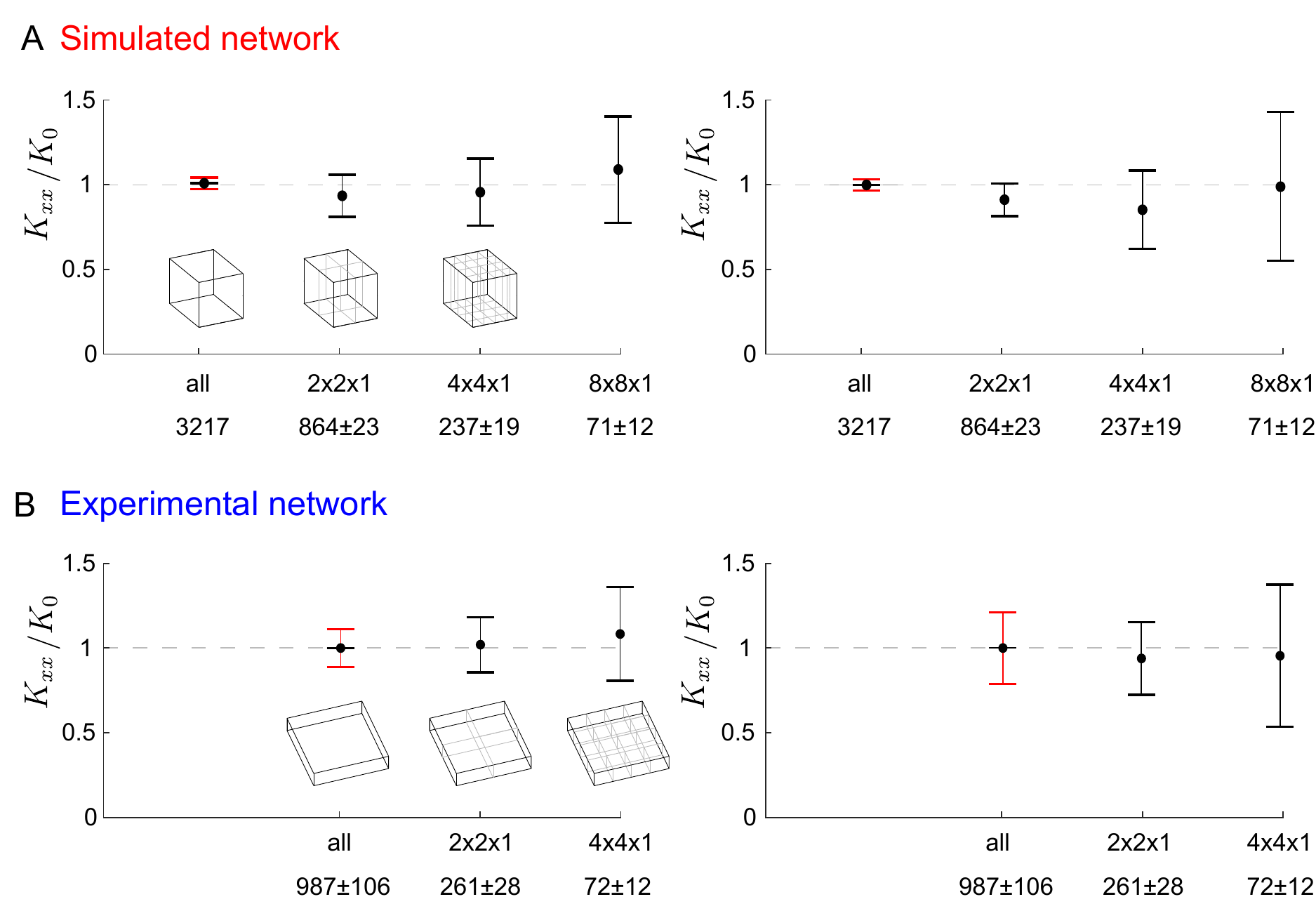}
\end{center}
\vspace{-5mm}
\caption[]{
\textbf{Effect of sample volume on computed network permeability.}
\textbf{A.}
Errorbars show variability of computed permeability of simulated networks
for entire sample volume ($1\times 1\times 1$, $n=3$ simulated networks),
sample volume divided in $4$ equal volumes ($2\times 2\times 1$, $m=4$ non-overlapping sub-samples for each of $n$ full simulated networks),
sample volume divided in $8$ equal volumes ($4\times 4\times 1$, $m=16$),
sample volume divided in $64$ equal volumes ($8\times 8\times 1$, $m=64$).
For the entire sample volume,
errorbars (red) denote the coefficient of variation of permeabilities of the $n$ full networks (standard deviation normalized by mean).
For the sub-sample volumes,
errorbars (black) denote the standard deviation of the pooled permeabilities of sub-samples ($n\cdot m$), 
where each permeability was normalized by the permeability of the corresponding full networks;
\textit{left:} permeability $K_{xx}$ along direction of nematic alignment,
\textit{right:} permeability $K_{yy}$ along $y$ axis perpendicular to direction of nematic alignment.
Number of edges is indicated below horizontal axes (mean$\pm$s.e.).
\textbf{B.}
Same as panel A, but for sinusoidal networks:
for entire sample volume ($1\times 1\times 1$, $n=3$ biological networks),
sample volume divided in $4$ equal volumes ($2\times 2\times 1$, $m=4$ non-overlapping sub-samples for each of $n$ biological networks),
sample volume divided in $8$ equal volumes ($4\times 4\times 1$, $m=16$).
}
\label{figureS1}
\vspace{-2mm}
\end{figure}

\vspace{3mm}
\textit{Network permeability computation biased for small sample volumes.}
Network permeability is an intensive measure that should be independent of network size for statistically homogeneous networks.
This property is the basis for effective medium theories.
In reality, however, small sample volumes can introduce a systematic bias in the computation of network permeabilities, see Fig.~\ref{figureS1}.
This provides additional motivation of our use of large network samples, 
as well as for our network generation algorithm to simulate spatially unrestricted networks with prescribed statistical properties.

\begin{figure}
\begin{center}
\includegraphics[width=15cm]{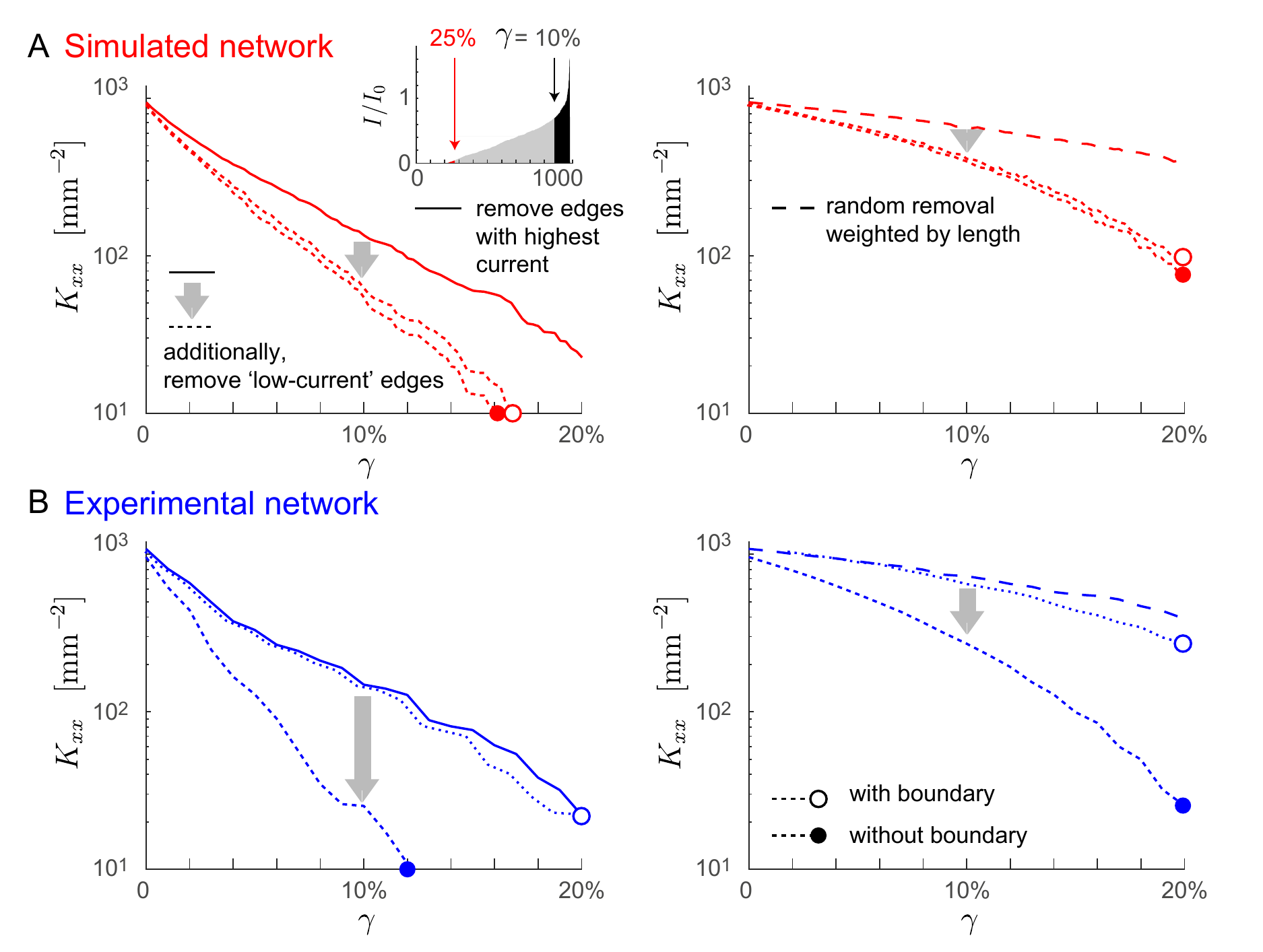}
\end{center}
\vspace{-5mm}
\caption[]{
\textbf{Low-current edges contribute to resilience.}
We repeated the analysis of Fig.~\ref{figure4}C, but additionally removed low-current edges
from both simulated and sinusoidal networks.
\textbf{A.}
\textit{Left:} 
We show again the permeability-at-risk curve for simulated networks from Fig.~\ref{figure4}C
(solid red, scenario i; $\lambda=0.2$, mean of $n=3$ simulated networks),
where a fraction $\gamma$ of edges is removed, 
starting with the edges that carry the highest current in the unperturbed network.
We compare this curve to a second permeability-at-risk curve, 
where additionally $25\%$ of the edges that carry the lowest current in the unperturbed network are removed 
(dotted, marked with open circle).
Alternatively, 
we can likewise remove $25\%$ of all edges,
choosing again those that carry the lowest current in the unperturbed network, 
yet restrict the selection to edges that do not touch the boundary of the simulation domain (dotted, marked with filled circle).
The fraction $\gamma$ is always measured relative to the total number of edges in the unperturbed network.
Hence, for the dotted curves, 
the number of edges of the perturbed networks equals $75\%-\gamma$ of the number of edges in the unperturbed networks.
\textit{Right:}
Same as left panel, but for random removal of a fraction $\gamma$ of edges (scenario ii), 
with removal probability proportional to edge length. 
\textit{B.}
Same as panel A, but for the sinusoidal networks.
Of note, the permeability-at-risk curves for the networks with low-current edges removed 
differ depending on whether we restrict the selection to low-current edges that do not touch the boundary 
(dotted curve, marked with filled circle), or not (dotted curve, marked with open circle).
This can be attributed to the small dimensions of the tissue sample, notably in $z$ direction, 
which results in a number of edges at the boundary that carry almost zero current.
}
\label{figureS2}
\vspace{-2mm}
\end{figure}

\vspace{3mm}
\textit{Permeability-at-risk if low-current edges removed.}
We repeated the analysis of Fig.~\ref{figure4}C, but additionally removed low-current edges.
If the networks were not perturbed otherwise, we find that their permeability almost did not change. 
However, if together with the low-current edges also a fraction $\gamma$ of edges is removed,
the permeability is substantially reduced, see Fig.~\ref{figureS2}.
This shows that low-current edges partially contribute to the resilience of networks, 
allowing for limited re-routing of flow if high-current edges are removed.

\vspace{3mm}
\textit{Minimal model of metabolic uptake.}
We consider a minimal model of metabolic uptake from the sinusoidal network by surrounding tissue
(for refined models, see e.g.\ Ref.\cite{Meigel2018}).
Let $c(\q,t)$ be the concentration of a molecule in solution in the flowing blood at time $t$ at position $\q$ in the sinusoidal network.
The dynamics of the concentration profile $c(\q,t)$ in the network is governed by diffusion, convection, and uptake
\begin{equation}
\frac{d}{dt} c(x,t) = -\nabla\cdot ( \v c ) \,+\, D \nabla^2 c \,-\, \beta c .
\end{equation}
Here, we assumed for simplicity a constant uptake rate $\beta$ with units of an inverse time.
Additionally, we assume that concentration is homogeneous across the cross-section of a sinusoid with constant area $A$. 
Thus, we need to consider only concentration profiles of a one-dimensional coordinate along a sinusoid edge.
Correspondingly, the flow velocity $\v$ for an edge with unit vector $\e$
can be taken as $\v = I/A \e$, where $I$ is the current through the edge.

For efficient computation on a network, 
we employ an upwind Euler scheme with concentrations evaluated at the node positions $\q_j$
with fixed time step $\Delta t$
\begin{equation}
\label{eq:metabolic_uptake}
c(\q_j, t+\Delta t) = 
- \sum_k \max\{0,I_{k,j}\} \, \left[ c(\q_k,t) - c(\q_j,t) \right]\, \frac{\Delta t}{V_j} % convection
\,+\, \sum_k D \left[ c(\q_k,t) - c(\q_j,t) \right] \, \frac{A \Delta t}{l_{j,k} V_j} % diffusion
\,-\, \beta c(\q_j,t) . % uptake
\end{equation}
Here, the sum $\sum_k$ extends over all nodes connected to node $j$,
with $l_{j,k}$ denoting the length of the edge connecting both nodes.
For each node $j$, we have introduced an \textit{associated volume} 
$V_j = \sum_k l_{j,k} A / 2$.
Note that due to Eq.~(\ref{eq:Kirchhoff}), 
these update rules conserve total mass $\sum c(\q_j)V_j$ if $\beta=0$.
To increase the spatial resolution, we divided each edge of the original network (see Fig.~\ref{figure3}) into $n=4$ equal parts.
For nodes $j$ with small $V_j$, 
Eq.~(\ref{eq:metabolic_uptake}) is modified by directly using the analytical solution for diffusive exchange between a pair of neighboring nodes,
which speeds up computations by allowing to choose larger time steps $\Delta t$.

\begin{figure}
\begin{center}
\includegraphics[width=15cm]{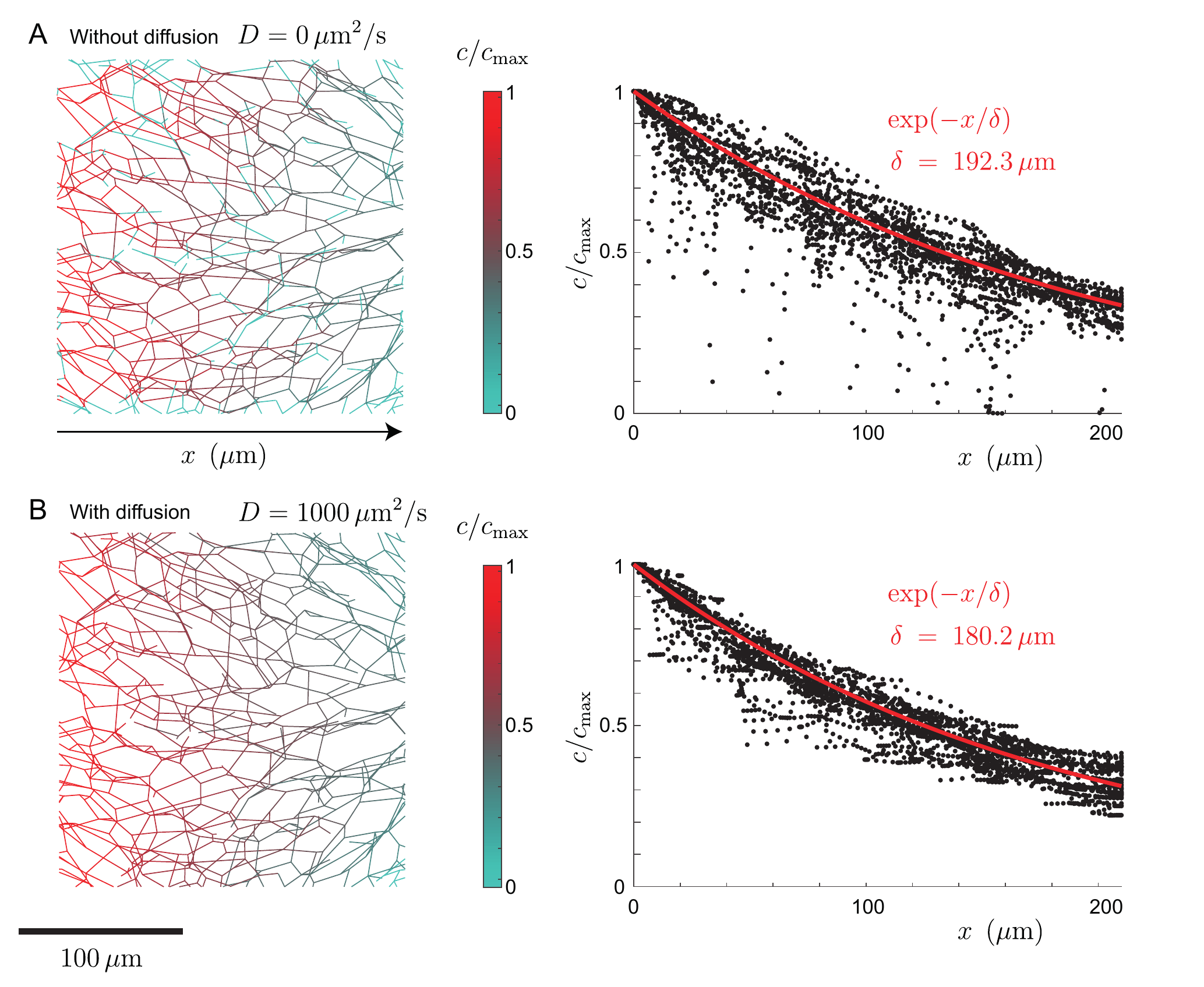}
\end{center}
\vspace{-5mm}
\caption[]{
\textbf{
Computed concentration profile of a prototypical metabolite using a minimal model of metabolic uptake from the sinusoidal network.
}
\textbf{A.} 
Steady-state concentration profile computed according to Eq.~(\ref{eq:metabolic_uptake})
for a case without diffusion. 
Approximately, the concentration profile decays exponentially along the pressure gradient direction
(black: steady-state concentration $c(\q_j)$ normalized by maximal concentration $c_\mathrm{max}$ as function of position $x=\q_j\cdot\e_x$,
red: fit $\exp(-x/\delta)$).
\textbf{B.} Same as panel A, but accounting for diffusion.
Parameters:
uptake rate $\beta=0.1\,\mathrm{s}^{-1}$,
diffusion coefficient $D=0$ in A, $D=1000\,\micron^2/\mathrm{s}$ 
($\sim$ diffusion coefficient of oxygen) in B;
for flow computation:
sinusoid radius $R=3\,\micron$, 
sinusoid cross-sectional area $A=\pi R^2$, 
dynamic viscosity of blood $\mu = 3.5 \cdot 10^{-3}\,\mathrm{Pa\,s}$ \cite{White2016},
resistance per unit length $\kappa = 8 \mu / (\pi R^4)$ (assuming Poiseuille flow),
pressure difference $\Delta p=\Delta p_0 L_x / L_0$ 
with
pressure difference between PV and CV $\Delta p_0 = 100\,\mathrm{Pa}$ \cite{Debbaut2012,Ricken2015},
typical distance between PV and CV $L_0 = 500\,\micron$, 
spatial dimension of region of interest $L_x = 210\,\micron$.
}
\label{figureS3}
\vspace{-2mm}
\end{figure}

Fig.~\ref{figure3} shows steady-state concentration profiles for the network flow from Fig.~\ref{figure3}D.
Remarkably, the spatial concentration profile varies homogeneously, 
despite the rather inhomogeneous distribution of flows in the network. 
Concentration decays approximately exponential along the flow direction;
the decay length $\delta\sim v/\beta$ 
is set by the ratio of a typical flow speed $v$ and uptake rate $\beta$.
Accounting for diffusion with $D>0$, homogenizes the concentration profile even further, see Fig.~\ref{figureS3}B.

\end{document}